\documentclass[aps,twocolumn,floatfix,superscriptaddress]{revtex4-1}
\usepackage{amsmath,amssymb,graphicx,cases}
\usepackage{epsfig}
\usepackage{textcomp}
\usepackage{epstopdf}
\usepackage{float}
\usepackage{braket}
\usepackage[normalem]{ulem}

\usepackage[normalem]{ulem}
\newcommand{\stkout}[1]{\ifmmode\text{\sout{\ensuremath{#1}}}\else\sout{#1}\fi}

\usepackage{color}
\usepackage[dvipsnames]{xcolor}
\definecolor{Blue}{rgb}{0.00, 0.00, 1.00}
\definecolor{Red}{rgb}{1.00, 0.00, 0.00}
\definecolor{Green}{rgb}{0.00, 0.50, 0.00}

\newcommand{\nn}{\nonumber}
\newcommand{\be}{\begin{equation}}
\newcommand{\ee}{\end{equation}}
\newcommand{\bea}{\begin{eqnarray}}
\newcommand{\eea}{\end{eqnarray}}

\renewcommand{\mod}{\text{ mod }}

\usepackage[colorlinks=true, urlcolor=blue, anchorcolor=blue, citecolor=blue,filecolor=blue,linkcolor=blue,menucolor=blue]{hyperref}

\usepackage{fixmath}
\newcommand{\vect}[1]{\mathbold {#1}} 

\begin{document}
\title{Large deviations in chaotic systems: exact results and dynamical phase transition}

\author{Naftali R. Smith}
\email{naftalismith@gmail.com}
\affiliation{Department of Solar Energy and Environmental Physics, Blaustein Institutes for Desert Research, Ben-Gurion University of the Negev,
Sede Boqer Campus, 8499000, Israel}


\begin{abstract}

Large deviations in chaotic dynamics have potentially significant and dramatic consequences.
We study large deviations of series of finite lengths $N$ generated by chaotic maps. The distributions generally display an exponential decay with $N$, associated with large-deviation (rate) functions.
We obtain the exact rate functions analytically for the doubling, tent, and logistic maps. 
For the latter two, the solution is given as a power series whose coefficients can be systematically calculated to any order.
We also obtain the rate function for the cat map numerically, uncovering strong evidence for the existence of a remarkable singularity of it that we interpret as a second order dynamical phase transition.
Furthermore, we develop a numerical tool for efficiently simulating atypical realizations of sequences if the chaotic map is not invertible, and we apply it to the tent and logistic maps.

\end{abstract}

\maketitle

\textit{Introduction.}--- 
Classical chaos is a fundamental property of a host of natural systems. 
It describes the unpredictability of deterministic dynamical systems, due to exponential growth of uncertainties in the initial conditions -- the ``butterfly effect''.
This notion gives rise to effective statistical descriptions of chaotic systems that form the foundations of statistical mechanics, for example, in terms of standard or anomalous diffusion \cite{Geisel82, Schell82, Fujisaka82, Sato19, AMR22}.
Rare events (large deviations) in chaotic dynamics can be extremely important, as they can have significant and potentially catastrophic consequences. One important example is extreme weather events such as heatwaves or floods \cite{WB16, RWB17, RB21, GLRW21, GL21, chaosWeather21, GLRW21, EPYTNLF22},  whose probabilities are especially challenging to predict under today's changing climate conditions which preclude using only historical data to assess their likelihood.
Additional examples are found in dynamics of stock markets \cite{KSB17}, road traffic \cite{STSAH02}, populations \cite{FBBC19, FBBB20, PAF20} and pandemics \cite{PAT21}.

However, while large deviations in stochastic systems have been extensively studied, both theoretically and numerically \cite{Varadhan,O1989,DZ,Hollander,Bray, Majumdar2007, T2009,Derrida11, RednerMeerson, bertini2015, Vivo2015, MeersonAssaf2017, MS2017,Touchette2018, Derrida2007, Waclaw2010,CohenMukamel2012, EvansMajumdar2014, Baek15, Baek17, Baek18,  Shpielberg2016, TouchetteMinimalModel, LeDoussal2017, SKM2018, Hartmann2002, bucklew2004,touchette2011},
large deviations in deterministic, chaotic systems have received somewhat less attention, see however 
\cite{WB16, RWB17, RB21, GLRW21, GL21, chaosWeather21, GLRW21, EPYTNLF22, FBBB20,
Young1990, Haller01, CLMPV02, Haller05,EP86, PV87, Frisch95, Bec06, AV15, LNKT13, BMV14, LLA14, JM15, PLP16, LLA17, DSDLV18, YYSL21}.
In particular, for chaotic systems there are fewer existing exact analytic results for the rate (large-deviation) function, which is a central object in the study of large deviations (see definition below). 
Of particular interest are dynamical phase transitions (DPTs): Singularities of rate functions that lead to distribution tails that are much larger or much smaller than one would naively expect.

 In principle, one would not expect there to be a fundamental difference between the large-deviation behaviors of chaotic and stochastic systems, because symbolic dynamics maps chaotic systems to stochastic ones. 
However, 
it is usually not straightforward to give explicit symbolic
dynamics, so it is not always easy to apply methods that work for stochastic systems to chaotic ones.
In numerical Monte Carlo (MC) simulations, these difficulties are usually circumvented by adding a weak noise term \cite{ACV17, WB16, RWB17}, although alternatives exist \cite{Grassberger85}.

Let us define the class of problems that we study here.
We consider a sequence $\vect{x}_1, \dots, \vect{x}_N$ of elements of $\mathbb{R}^d$ generated by a chaotic map $f(\vect{x})$ via
$\vect{x}_{i+1} = f(\vect{x}_i)$,
where $\vect{x}_1$ is randomly sampled from the invariant measure (IM) of the process $p_s(\vect{x})$.
We recall that the IM is the measure that is preserved by the map $f(\vect{x})$, i.e., if $\vect{x}$ is distributed according to the IM, so is $f(\vect{x})$.
We quantify 
fluctuations in the system by studying the full distribution of dynamical observables
\be
\label{Adef}
A=\frac{1}{N}\sum_{i=1}^{N}g\left(\vect{x}_{i}\right) \,,
\ee
where $g:\mathbb{R}^d \to \mathbb{R}$. 
The study of dynamical observables has been an important theme in the ongoing research of large deviations, and importantly for the following, it is amenable to theoretical analysis via the powerful Donsker-Varadhan (DV) formalism \cite{Varadhan, Bray, T2009,MS2017, Touchette2018}. The DV theory is more commonly formulated for stochastic systems, however its formulation for chaotic systems is straightforward and has been known for some time \cite{GBP88, ParisiAppendix84, AV15}.
Note that $A$ is a deterministic function of the initial condition $A=\mathcal{A}\left(\vect{x}_{1}\right)$.
In the large-$N$ limit, for ergodic dynamics, $A$ converges to its ensemble-average value
$A\to\int g\left(\vect{x}\right)\,p_{s}\left(\vect{x}\right)d\vect{x}\equiv a_{*}$
with probability 1.
Individual realizations, however, deviate from this value due to fluctuations in the initial condition $\vect{x}_1$. 
For the particular case $g(\vect{x}) = \ln |J_f(\vect{x})|$, where $J_f$ is the Jacobian determinant \cite{BS93, Anteneodo04}, $A$ corresponds to the finite-time Lyapunov exponent \cite{GBP88, Haller01, CLMPV02, Haller05,EP86, PV87, Frisch95, Bec06, LNKT13, BMV14, LLA14, JM15, PLP16, LLA17, DSDLV18, YYSL21},
which describes the rate of separation with respect to nearby trajectories over a finite time.

The DV 
framework predicts that, under fairly general conditions, fluctuations obey a large-deviation principle (LDP) \cite{GBP88, ParisiAppendix84, T2009, AV15, footnote:PDF}
\be
\label{DVScaling}
P\left(A=a\right)\sim e^{-NI\left(a\right)}, \quad N \to \infty,
\ee
where $I\left(a\right)=-\lim_{N\to\infty}\frac{\ln P\left(A=a\right)}{N}$, the ``rate function'', encodes the system's dynamical behavior. $I(a)$ is convex and vanishes at its minimum, which is at $a = a_*$.
Note that in Markov chains (sequences generated by stochastic maps), the initial condition is generally not important because it is quickly ``forgotten'' as a result of the randomness of the dynamics.
In contrast, for a deterministic, chaotic system, \emph{all} of the randomness enters in the initial condition.
As has been known for quite some time \cite{GBP88, ParisiAppendix84, T2009, AV15},
$I(a)$ can be calculated by solving an auxiliary problem of finding the largest eigenvalue of a ``tilted operator'' which is related to the generator of the dynamics, which in the present case is the Frobenius-Perron operator.
In this Letter, we carry out this calculation explicitly and obtain the exact rate functions $I(a)$ for the doubling, tent and logistic maps for particular observables \cite{footnote:mappings}.
For the cat map, we compute $I(a)$ numerically, uncovering strong evidence for the existence of a remarkable singularity of it that we interpret as a DPT, which signals a sudden change in the way that the system realizes a given value $A=a$ as $a$ crosses the critical point.
Furthermore, for noninvertible maps $f(\vect{x})$, we develop a MC algorithm that efficiently samples realizations that reach unlikely values of $A$ by generating the sequence $\vect{x}_1,\dots,\vect{x}_N$ in reverse order and in a biased manner.

\textit{Theoretical framework.}---
The theoretical framework that we use to obtain the scaling \eqref{DVScaling} and calculate $I(a)$ has been known (in various forms) for decades \cite{GBP88, ParisiAppendix84, T2009, AV15}, and we recall it here for the sake of completeness.
It is useful to first consider the scaled cumulant generating function (SCGF) $\lambda(k)$, defined as
$\left\langle e^{NkA}\right\rangle \sim e^{N\lambda\left(k\right)}$. $\lambda(k)$ is found by calculating the largest eigenvalue of a ``tilted'' (modified) generator of the dynamics, where $k \in \mathbb{R}$ is the tilting parameter. The existence of a nonzero $\lambda(k)$ yields the scaling \eqref{DVScaling}, 
and $I(a)$ is then obtained via a Legendre-Fenchel transform \cite{GBP88, ParisiAppendix84, T2009}
$I(a) = \sup_{k \in \mathbb{R}} \left[ k a - \lambda(k) \right]$.

We first note that if $\rho(\vect{x})$ is the probability distribution function (PDF) of some element 
$\vect{x}=\vect{x}_i$,
then the PDF of the next element $\vect{y}=\vect{x}_{i+1}$ is $L\rho\left(\vect{y}\right)$, where $L$ is the Frobenius-Perron operator,
\be
\! L\rho\left(\vect{y}\right) \! = \! \int \! \rho\left(\vect{x}\right)\delta\left(\vect{y}-f\left(\vect{x}\right)\right)d\vect{x} \! = \!\! \sum_{\vect{z}=f^{-1}\left(\vect{y}\right)}\frac{\rho\left(\vect{z}\right)}{ \left|J_{f}\left(\vect{z}\right)\right|}  .
\ee
The SCGF $\lambda(k)$ is equal to the logarithm of the largest (real) eigenvalue of the ``tilted'' operator \cite{Ellis, DZ, GBP88, ParisiAppendix84, T2009, SM}
\be
\label{Lkdef}
L_{k}\rho\!\left(\vect{y}\right)\!=\!\!\int\!e^{kg\left(\vect{x}\right)}\delta \! \left(\vect{y}-f\!\left(\vect{x}\right)\right) \! \rho\!\left(\vect{x}\right)d\vect{x}\!=\!\!\!\sum_{\vect{z}\in f^{-1}\left(\vect{y}\right)}\!\!\!\frac{e^{kg\left(\vect{z}\right)} \! \rho\!\left(\vect{z}\right)}{\left|J_{f}\left(\vect{z}\right)\right|}.
\ee
Note that $L_{k=0} = L$ whose largest eigenvalue is $e^{\lambda(k=0)}=1$, the eigenvector being the IM $p_s(\vect{x})$.
It is convenient to define $\psi\left(\vect{x}\right)=e^{kg\left(\vect{x}\right)}\rho\left(\vect{x}\right)$, so the equation $L_k\rho\left(\vect{x}\right)=e^{\lambda\left(k\right)}\rho\left(\vect{x}\right)$ becomes
\be
\label{Ltilde}
\tilde{L}_{k}\psi\left(\vect{x}\right)=e^{kg\left(\vect{x}\right)}\sum_{\vect{z}\in f^{-1}\left(\vect{x}\right)}\frac{\psi\left(\vect{z}\right)}{ \left|J_{f}\left(\vect{z}\right)\right|}=e^{\lambda\left(k\right)}\psi\left(\vect{x}\right)\,.
\ee
The calculation of the full distribution of $A$ is thus mapped to the auxiliary problem of calculating the largest eigenvalue $\lambda(k)$ of the operator $\tilde{L}_{k}$ \cite{GBP88, ParisiAppendix84, T2009, AV15}.
We emphasize that $\lambda(k)$ depends on the observable in question through the function $g(\vect{x})$ which enters in Eq.~\eqref{Ltilde}. As a result, the rate function $I(a)$ too depends on $g(\vect{x})$.

\textit{MC algorithm.}---
We now describe our numerical algorithm for the efficient MC simulation of unusual values of $A$ for noninvertible maps $f(\vect{x})$.
An alternative, statistically equivalent 
method for generating random sequence realizations, in the form of a Markov chain, is by first randomly sampling $\vect{x}_N$ from the IM $p_s(\vect{x})$ and then \emph{stochastically} generating the elements of the sequence in reverse order by choosing $\vect{x}_i$ from the set $\vect{z}\in f^{-1}\left(\vect{x}_{i+1}\right)$ 
with probabilities $p_{s}\left(\vect{z}\right)/\left[p_{s}\left(\vect{x}_{i+1}\right)\left|J_{f}\left(\vect{z}\right)\right|\right]$ \cite{SM}.
Such reverse simulations have been employed successfully before, see e.g. \cite{Grassberger85}. Importantly, they involve stochasticity, and therefore 
they enable one to \emph{bias} the simulations, by choosing from among the preimages with probabilities that are different to those given above \cite{Grassberger85}, a principle that we exploit in order to 
bias our simulations toward atypical values of $A$. 
Let us demonstrate this by considering the particularly simple case of $d=1$, and assume that every $x$ has exactly two preimages $z_1, z_2$, with $p_{s}\left(z_{1}\right)/\left|f'\left(z_{1}\right)\right| = p_{s}\left(z_{2}\right)/\left|f'\left(z_{2}\right)\right|$, as is the case for each of the doubling, tent and logistic maps considered below.
We define $N$ indicator random variables $\xi_{1}, \dots, \xi_{N}$, where  $\xi_i = 1$ (0) if $x_i$ is the larger (smaller) of the two preimages of $f(x_{i})$. 
From the definition of the stochastic reverse process, the $\xi_i$'s are independent and identically distributed Bernoulli random variables with $\mathbb{P}\left(\xi_{i}=0\right)=\mathbb{P}\left(\xi_{i}=1\right)=1/2$, and as a result, their sum $B=\sum_{i=1}^{N}\xi_{i}$ is binomially distributed, $\mathbb{P}\left(B=b\right)=\left( \! \begin{array}{c}
N\\
b
\end{array} \! \right)2^{-N}$.
Using the law of total probability, we have
\be
\label{totalProbability}
P\left(A = a\right)=\sum_{b=0}^{N}P\left(A = a \, | \, B=b\right)\mathbb{P}\left(B=b\right) \, .
\ee 
The reverse process can be simulated conditioned on $B$ taking a specified value $b$, by (i) randomly choosing a  subset $\mathcal{I} \subset\left\{ 1,\dots,N\right\} $ of size $b$ 
[each subset is chosen with the same probability $\left(\!\begin{array}{c}
N\\
b
\end{array}\!\right)^{-1}$],
(ii) randomly sampling a number $x_{N+1}$ from the IM, and
(iii) calculating $x_1,\dots,x_{N}$ in reverse order, where $x_i$ is given by the larger (smaller) preimage of $x_{i+1}$ if $i \in \mathcal{I}$ ($i \notin \mathcal{I}$).
One then computes $P\left(A=a \, |\, B=b\right)$ from MC simulations of these restricted dynamics. 
Repeating this process for $b=0,1,\dots,N$, one then computes $P(A=a)$ from Eq. \eqref{totalProbability}.
Atypical values of $A$ are thus accessed, if they tend to occur concurrently with atypical values of $B$.

\textit{Applications.}---
We now apply these tools to study several standard chaotic maps, beginning with the doubling map
$f\left(x\right) =2x\mod1$, where $x\in[0,1]$ and $z\mod1$ is the fractional part of $z$, with $g(x) = x$. Here, $f^{-1}\left(x\right)=\left\{ x/2,\left(x+1\right)/2\right\} $, so Eq.~\eqref{Ltilde} reads
\be
\frac{e^{kx}}{2}\left[\psi\left(\frac{x}{2}\right)+\psi\left(\frac{x+1}{2}\right)\right]=e^{\lambda\left(k\right)}\psi\left(x\right)\,,
\ee
whose solution,
\be
\label{lambdaDoubling}
\psi\left(x\right)=e^{2kx},\quad\lambda\left(k\right)=\ln\left(\left(1+e^{k}\right)/2\right)\,,
\ee
gives the rate function through a Legendre transform \cite{footnote:Legendre, LegendreNutshell},
\be
\label{IaDoubling}
I(a) = a\ln a+\left(1-a\right)\ln\left(1-a\right)+\ln2 \, .
\ee
As the reader may have noticed, $I(a)$ is precisely the rate function that describes a binomial distribution (see, e.g., \cite{MS2017}). 
We explain this coincidence in \cite{SM} by calculating $I(a)$ via an alternative method, providing a validation of \eqref{IaDoubling}.

\begin{figure*}[t]
\includegraphics[width=0.47\linewidth,clip=]{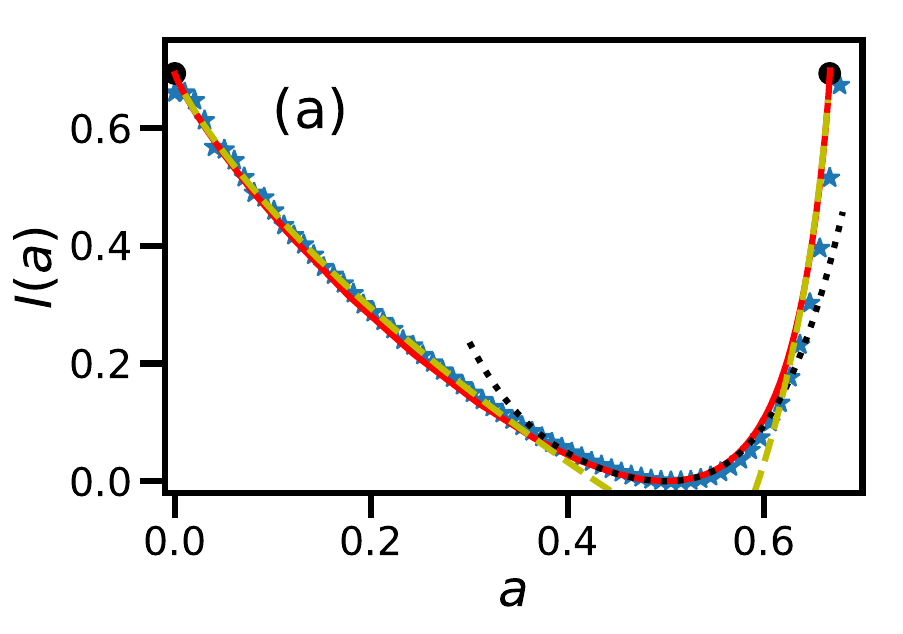}
\hspace{2mm}
\includegraphics[width=0.47\linewidth,clip=]{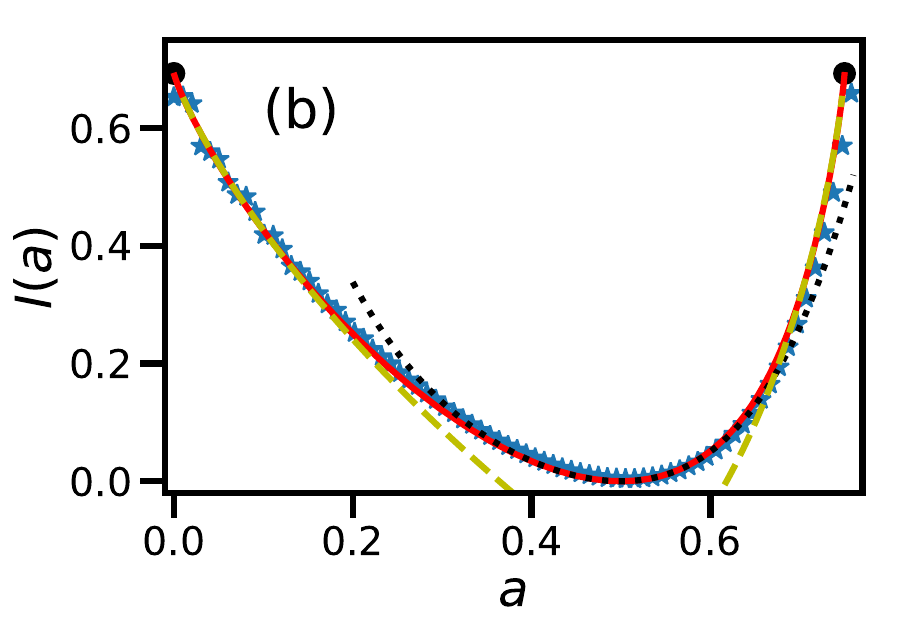}
\caption{Rate functions $I(a)$ that describe the full distributions $P(A=a)$ at $N\gg1$ for the tent map (a) and logistic map (b).
Solid lines were obtained through numerical diagonalizations of $\tilde{L}_k$ from Eqs.~\eqref{psiTentEq} and \eqref{psiLogisticEq}, respectively. Dotted lines are the exact analytic solutions, evaluated up to order $O\left(\left(a-1/2\right)^{4}\right)$, Eqs.~\eqref{IaTent} and \eqref{IaLogistic}, respectively. 
Markers correspond to properly rescaled data from biased MC simulations with $N=50$ (see \cite{SM} for details).
The points $I(0)=I(2/3)=\ln2$ in (a) and $I(0)=I(3/4)=\ln2$ in (b) are marked by $\bullet$.
Dashed lines correspond to the asymptotic behaviors of the rate functions near the edges of their supports, that we calculate in \cite{SM}.}
\label{figIaTentLogistic}
\end{figure*}

Let us now consider the tent map,
$f(x) = 1 - |1-2x|$,
again with $g(x) = x$.
Before turning to the calculation of $I(a)$, we make two observations: (i) As we find in \cite{SM}, $A$ is strictly bounded from above by $m_{N}=\max_{x_{1}\in\left[0,1\right]}\mathcal{A}\left(x_{1}\right)$ which, at $N \to \infty $, converges to $m_{\infty}=2/3$, the nontrivial fixed point of $f(x)$.
We therefore anticipate that the support of $I(a)$ is the interval $[0,2/3]$, which, as shown below, is indeed the case.
This is nontrivial, since the IM for the tent map is uniform over the entire interval $[0,1)$.
(ii) If $x_1\sim2^{-N}$, then $A\sim 1/N$. The probability for this is $\sim2^{-N}$, which, using  \eqref{DVScaling}, leads to the bound $I(0) \le \ln 2$.
Similarly, if $\left|x_{1}-2/3\right|\sim2^{-N}$ then $\left|A-2/3\right|\sim1 / N$, leading to $I(2/3) \le \ln 2$.
In fact, we find below that these inequalities are saturated, $I\left(0\right)=I\left(2/3\right)=\ln2$.

Let us calculate $I(a)$.
For the tent map, Eq.~\eqref{Ltilde} reads
\be
\label{psiTentEq}
\! \tilde{L}_k\psi\left(x\right) = \frac{e^{kx}}{2} \! \left[\psi\left(\frac{x}{2}\right)+\psi\left(1-\frac{x}{2}\right)\right] \! =e^{\lambda\left(k\right)}\psi\left(x\right) .
\ee
We now present an exact solution to Eq.~\eqref{psiTentEq}, which we obtain in the form of a perturbation theory in $k$ that can be exactly solved at all orders. 
The IM $p_s(x)$ is uniform over $x \in [0,1)$, and indeed, one finds that for $k=0$, $\psi\left(x\right)=1$ and $\lambda(k)=0$.
We expand in $k$,
\bea
\psi \left(x\right) &  = & 1+k\psi_{1}  \left(x\right)+k^{2}\psi_{2}  \left(x\right)+\dots, \\
\lambda \left(k\right) & = & k\lambda_{1}+k^{2}\lambda_{2}+\dots.
\eea
Let us first find the solution to first order in $k$.
Keeping terms up to order $O(k)$ in Eq.~\eqref{psiTentEq}, we obtain
\be
\label{psiTentEq1}
1+\frac{k}{2}\left[2x+\psi_{1}\left(\frac{x}{2}\right)+\psi_{1}\left(1-\frac{x}{2}\right)\right]\!=\!1+k\left[\lambda_{1}+\psi_{1}\left(x\right)\right]
\ee
whose exact solution is 
$\psi_{1}\left(x\right)=x,\;\lambda_{1}=1/2 .$
This perturbative procedure can be explicitly carried out to arbitrary order in $k$, yielding the exact rate function $I(a)$. $\psi_n(x)$ turns out to be a polynomial  of degree $n$, whose coefficients are found by solving a set of linear equations. In \cite{SM}, we work out the leading orders explicitly, and obtain
\be
\lambda\left(k\right)=\frac{k}{2}+\frac{k^{2}}{24}-\frac{k^{3}}{72}+\frac{41k^{4}}{8640}+\dots,
\ee
whose Legendre transform is
\be
\label{IaTent}
I\left(a\right)=6\left(a-\frac{1}{2}\right)^{2} \! +24\left(a-\frac{1}{2}\right)^{3} \! +\frac{588}{5}\left(a-\frac{1}{2}\right)^{4} \! +\dots.
\ee
 We give the solution up to eighth order in \cite{SM}.

We now consider the logistic map \cite{Feigenbaum78} at the Ulam point, $f\left(x\right)=4x\left(1-x\right)$, where $x\in[0,1)$, 
with $g(x)=x$.
The analysis is very similar to that of the tent map \cite{SM}.
This time, the support of $I(a)$ is $[0,3/4]$, $x=3/4$ being a fixed point of $f(x)$, with $I(0) = I(3/4) = \ln 2$, and Eq.~\eqref{Ltilde} reads
\be
\label{psiLogisticEq}
\tilde{L}_k\psi \! \left(x\right) \! = \! \frac{e^{kx}\left[\psi\left(\frac{1+\sqrt{1-x}}{2}\right)+\psi\left(\frac{1-\sqrt{1-x}}{2}\right)\right]}{4\sqrt{1-x}} \! = \! e^{\lambda\left(k\right)}\psi \! \left(x\right) .
\ee
Like in the tent map, we solve Eq.~\eqref{psiLogisticEq} perturbatively in $k$ to arbitrary order.
Expanding
\be
\psi\!\left(x\right)=p_{s}\left(x\right)\left[1+k\psi_{1}\!\left(x\right)+k^{2}\psi_{2}\!\left(x\right)+\dots\right]
\ee
where $p_{s}\left(x\right) = \left[\pi\sqrt{x\left(1-x\right)}\right]^{-1}$ is the IM \cite{Ulam47},
and $\lambda\!\left(k\right)=k\lambda_{1}+k^{2}\lambda_{2}+\dots$, we again find that $\psi_n(x)$ is a polynomial of degree $n$ whose coefficients can be found explicitly. 
In \cite{SM} we find 
\be
\lambda(k) = \frac{k}{2}+\frac{k^{2}}{16}-\frac{k^{3}}{64}+\frac{3k^{4}}{1024}+\dots, 
\ee
whose Legendre transform is
\be
\label{IaLogistic}
I\left(a\right)=4\left(a-\frac{1}{2}\right)^{2} \! + 8\left(a-\frac{1}{2}\right)^{3}\! +  24\left(a-\frac{1}{2}\right)^{4} \! +\dots.
\ee
The solution up to sixth order is found in \cite{SM}.
As shown in Fig.~\ref{figIaTentLogistic}, Eqs.~\eqref{IaTent} and \eqref{IaLogistic} are in excellent agreement with numerical computations of $P(A=a)$ from biased MC simulations with $N=50$, 
and with semi-analytic calculations of $I(a)$ obtained by
computing the largest eigenvalue $e^{\lambda(k)}$ of $\tilde{L}_{k}$ numerically using Ulam discretization \cite{KKS16}, and then performing the Legendre transform numerically.
Also plotted are the asymptotic behaviors of the rate functions near the edges of their supports, which we obtain in \cite{SM} by solving the eigenvalue problems in the limits $k \to \pm \infty$.
%
Before moving on, we note that for the doubling, tent and logistic maps, $I(a)=\ln 2$  -- which, in all these systems, equals the Lyapunov exponent -- at the edges of its support. We speculate that this feature may be universal for $d=1$.

We now turn to Arnold's cat map, where we uncover strong numerical evidence pointing at the existence of a remarkable DPT in $I(a)$. Here $d=2$, and $\vect{x}_i = \left(y_{i},z_{i}\right) \in \left[0,1\right]\times\left[0,1\right]$.
The cat map is defined by
\be
f\left(y,z\right)=\left(\left(y+z\right)\mod1,\left(y+2z\right)\mod1\right) \, .
\ee
Its IM is uniform on the unit square. We consider $g\left(y,z\right)=\left(y+z\right)/2$ \cite{footnote:Cat}.
$f(y,z)$ is invertible, with $\left|J_{f}\left(y,z\right)\right|=1$, so Eq.~\eqref{Ltilde} becomes 
\be
\label{psiCatEq}
e^{k\left(y+z\right) \! /2}\psi\!\left(\left(2y\! -\! z\right)\!\mod\! 1,\left(z\! -\! y\right)\!\mod\! 1\right)\! =\! e^{\lambda\left(k\right)}\psi\!\left(y,z\right).
\ee
Eq.~\eqref{psiCatEq} proved difficult to solve analytically or even numerically,
because of instabilities of the Ulam method (that occur even for $k \! = \! 0$ \cite{BSTV97, YYSL21}). 
Though we did not solve Eq.~\eqref{psiCatEq}, we are nevertheless able to predict some features of $I(a)$  by making the following observations:
(i) The dynamics are statistically invariant under the transformation $\left(y_{i},z_{i}\right)\to\left(1-y_{i},1-z_{i}\right)$.
Therefore, $P\left(A =a\right)=P\left(A = 1-a\right)$ is exactly symmetric, implying $I\left(a\right)=I\left(1-a\right)$. 
(ii) One can check that the joint distribution of any pair $\left(\xi_{1},\xi_{2}\right)$ of distinct elements taken from the set
$\left\{ y_{1},z_{1},\dots,y_{N,}z_{N}\right\} $
is uniform on the unit square, implying that $\xi_{1},\xi_{2}$ are statistically independent. Therefore, using $\left\langle y_{i}\right\rangle =\left\langle z_{i}\right\rangle =1/2$ and $\text{Var}\, y_{i}=\text{Var}\, z_{i}=1/12$, we find $\left\langle A\right\rangle =1/2$ and $\text{Var}A=1/(24N)$ so, using \eqref{DVScaling}, we anticipate the parabolic behavior
\be
\label{IaCat}
I\left(a\right)=12\left(a-1/2\right)^{2}+o\left(\left(a-1/2\right)^{2}\right) \, ,
\ee
corresponding to a Gaussian distribution of typical fluctuations as described by (an extension of) the central limit theorem.
(iii) The Lyapunov exponents of the cat map are $\pm 2 \ln \varphi$, where $\varphi=\left(1+\sqrt{5}\right) \! /2=1.618\dots$ is the golden ratio.
Therefore, if $y_1,z_1 \sim \varphi^{-2N}$, then the elements of the sequence grow with $i$ as $y_{i}, z_i \sim\varphi^{2i-2N}$, and as a result, $A \sim 1/N$.
The probability for this is $\sim \varphi^{-4N}$, leading to the bound $I\left(0\right)\le4\ln\varphi=1.9248\dots$ and similarly for $I(1)$.

\begin{figure*}[t]
\includegraphics[width=0.46\linewidth,clip=]{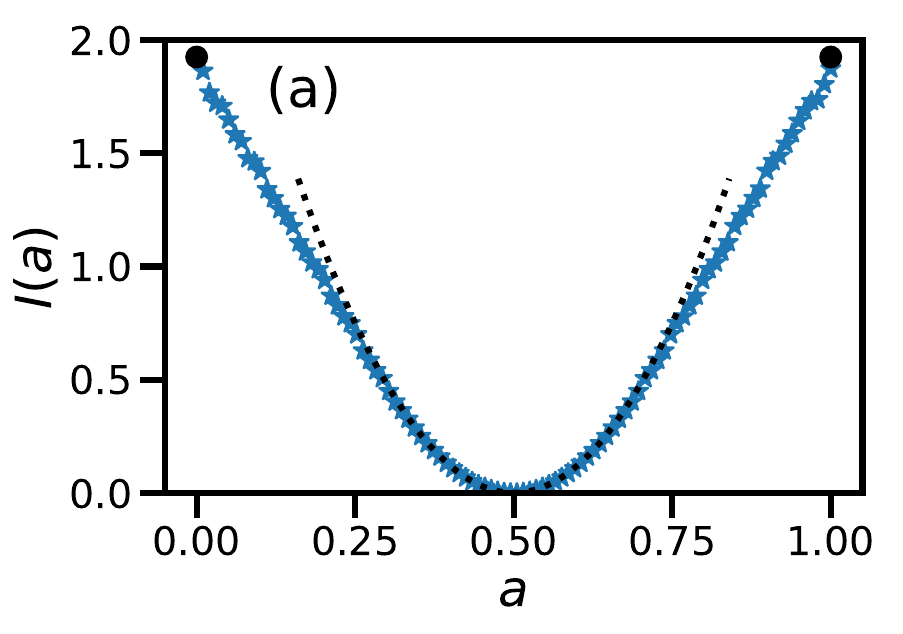}
\hspace{2mm}
\includegraphics[width=0.47\linewidth,clip=]{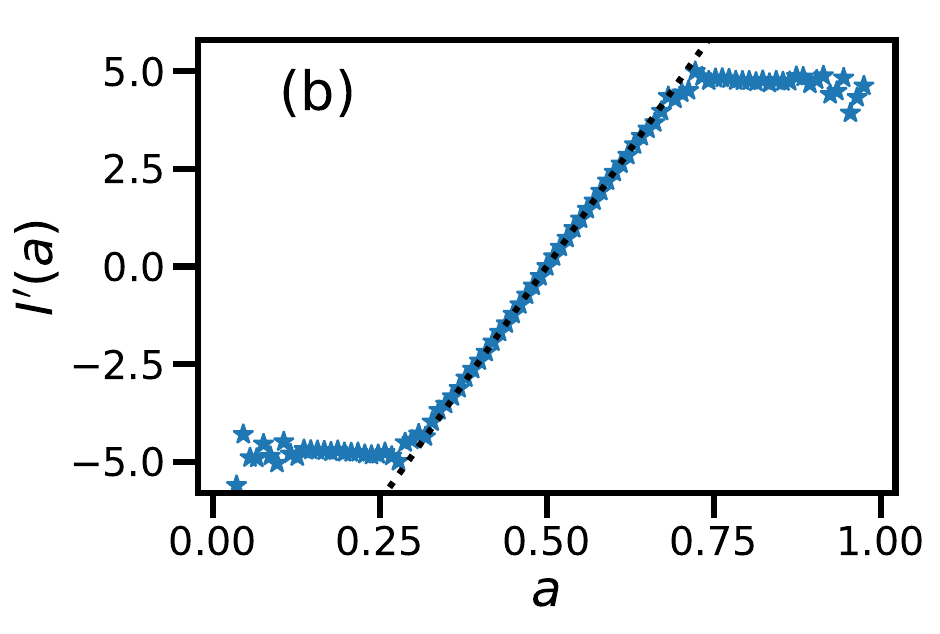}
\caption{(a) Markers correspond to data from $2 \times {10}^{10}$ direct MC simulations with $N=10$ for the cat map \cite{SM}, showing good agreement with the prediction \eqref{IaCat} (dotted line) at $a\simeq1/2$. The points $I(0)=I(1)=4\ln\varphi$ are marked by $\bullet$.
(b) $I'(a)$, showing a corner singularity (second-order DPT) at $a=a_c$ and $a = 1-a_c$, and taking constant values at $a\in\left[0,a_{c}\right]$ and at $a\in\left[1-a_{c},1\right]$.}
\label{figIaCat}
\end{figure*}

In Fig.~\ref{figIaCat}(a), we plot $I(a)$, which we computed from direct MC simulations with $N=10$ (since the cat map is invertible, we could not use the algorithm introduced in this work, and had to resort to direct MC simulations instead). 
Good agreement with the prediction of Eq.~\eqref{IaCat} is observed at $a \simeq 1/2$. In addition, we found that the bounds given above for $a=0$ and $a=1$ are in fact saturated, $I\left(0\right)=I\left(1\right)=4\ln\varphi$.

Far more remarkable, however, $I(a)$ appears to behave \emph{exactly} linearly for $a\in\left[0,a_{c}\right]$ with $a_c \simeq 0.3$ [Due to the symmetry $I(a)=I(1-a)$, this occurs symmetrically at $a \in [1-a_c,1]$ too). 
This is seen more clearly in Fig.~\ref{figIaCat}(b), where $I'(a)$  is plotted. Indeed, $I'(a)$  appears to have a corner singularity at $a = a_c$, and to take a constant value $I'(a) \simeq  -4.7$ for $a\in\left[0,a_{c}\right]$.
In the distribution \eqref{DVScaling}, $I(a)$ assumes the role of an effective free energy, and we therefore 
 interpret this singularity as a second-order DPT, because $I(a)$ and $I'(a)$ are continuous at the transition, but $I''(a)$ jumps \cite{footnote:2ndOrder}.
This (apparent) DPT constitutes a central result of this Letter.
At $a\in\left[0,a_{c}\right]$, we expect the system to display coexistence between the $a=0$ and $a=a_c$ states, meaning that for a fraction $a/a_c$ ($1 - a/a_c$) of the dynamics, the system will display the statistical behavior that corresponds to $a=0$ ($a=a_c$). 
In particular, we expect the distribution $P\left(x|A=a\right)$ of each element in the sequence, conditioned on observing a given $A = a\in\left[0,a_{c}\right]$, to be given by the superposition 
\be
P\left(x|A=a\right)=\left(1-\frac{a}{a_{c}}\right)P\left(x|A=0\right)+\frac{a}{a_{c}}P\left(x|A=a_{c}\right)
\ee
of the corresponding distributions conditioned on observing $A = 0$ and $A = a_c$, respectively.
This interpretation draws an analogy to DPTs that were found in stochastic systems, see e.g., \cite{Baek17, Baek18, TouchetteMinimalModel}.
%

%
%
\textit{Discussion.}--- We studied dynamical observables in the doubling, tent, logistic and cat maps. 
By using an existing theoretical framework \cite{GBP88, ParisiAppendix84, T2009, AV15},
we calculated the rate functions exactly (for particular observables) in the former three maps, where for the tent and logistic maps, our result is given in the form of a perturbation theory that can be solved to all orders.
Moreover, we calculated the rate function numerically for the cat map.
The rate functions $I(a)$ that we found have interesting properties: 
(i) For the tent and logistic maps, the rate functions are asymmetric although the IMs are symmetric, and in fact even the rate functions' supports differ from those of the IMs.
(ii) In all of the cases studied here, the supports of the rate functions are related to fixed points of the map $f(\vect{x})$, and at the edges of their supports, the rate functions take values that are related to the system's Lyapunov exponents. 
(iii) For the cat map, $I(a)$ has a remarkable singularity that we interpret as a second-order DPT, causing unusual values of $A$ to be far likelier than one would expect by extrapolating from the central part of the distribution.
It would be interesting to extend our results to other maps and/or to other observables.

As an alternative approach to ours, one could characterize the set of initial conditions $x_1$ for which $\mathcal{A}(x_1) = a$, since the statistical weight of this set (according to the IM) gives $P(A=a)$. In the limit $N \to \infty$, this set becomes 
\be
\label{Sa}
S_{a}=\left\{ x_{1}\,|\,\lim_{N\to\infty}\frac{1}{N}\sum_{n=0}^{N-1}g\left(f^{\left(n\right)}\left(x_{1}\right)\right)=a\right\} \,.
\ee
For instance, for $f(x) = $ the doubling map and $g(x)=x$, $S_{a}$ is the set of numbers $x_1 \in [0,1]$ whose binary representation has a ratio of $1-a:a$ between zeros and ones \cite{SM}.
One could then explore possible connections between our rate function $I(a)$ and various fractal dimensions of $S_a$, see \cite{OWY84, KP87, GBP88, ParisiAppendix84, BPPV84, Jensen85, Frisch85} where fractal dimensions were studied, and in particular, phase transitions were found \cite{OWY84, KP87, GBP88}. 

Finally, in some stochastic systems, the scaling \eqref{DVScaling} was recently found to break down and give way to anomalous scalings of large deviations \cite{NT18, MeersonGaussian19, GM19, Jack20, BKLP20,  MLMS21, MGM21, GIL21, GIL21b, Smith22OU, SM22} or of the distributions' cumulants \cite{Krajnik22a, Krajnik22b}. It would be interesting to search for anomalous scalings in deterministic, chaotic systems.

\bigskip

\textit{Acknowledgments---}
I thank Satya N. Majumdar and Grégory Schehr for pointing out useful references, Roiy Sayag and Tal Agranov for their comments on the manuscript, and Domenico Lippolis for a very helpful correspondence concerning eigenvalues of the Frobenius-Perron operator in the cat map.

\renewcommand{\theequation}{S\arabic{equation}}
\setcounter{equation}{0}
\renewcommand{\thefigure}{S\arabic{figure}}

\renewcommand*{\citenumfont}[1]{S#1}
\renewcommand*{\bibnumfmt}[1]{[S#1]}

\begin{widetext}
	\newpage

	\section*{Supplemental Material to the paper ``Large deviations of dynamical observables in chaotic systems and dynamical phase transition in the cat map" by N. R. Smith}

\subsection{Theoretical framework}
\label{sec:LDP}

Here, for completeness, we give some details regarding the theoretical framework that we use.
 This framework has been formulated long ago for chaotic maps \cite{GBP88s, ParisiAppendix84s,  AV15s}, but we rederive it here in the language of stochastic systems.
We first review the more general framework for Markov chains, in which large deviations have been studied both for continuous and discrete time using the Donsker-Varadhan formalism \cite{DeBacco2016s,TsobgniNyawo2016s,Whitelam2018s,CMT19s,Gutierrez2021s,Whitelam2021s, CVC22s, Monthus22s}.
A (discrete-time) Markov chain is a finite sequence $x_1, \dots, x_N$ with a finite state space, $x_{i}\in\left\{ 1,\dots,M\right\} $, such that the probability distribution of each element depends only on the previous one
\be
P\left(x_{i+1}=y\right)=\sum_{x=1}^{M}P\left(x_{i}=x\right)\Pi_{xy},
\ee
where $\Pi_{xy}$ is an $M\times M$ matrix of transition probabilities.
$x_1$ can be sampled from the steady-state distribution.
For ergodic Markov chains, the G\"{a}rtner-Ellis theorem states  \cite{GartnerElliss, DZs, HollanderSM, T2009s} that dynamical observables follow a large-deviations principle (LDP) at $N\to\infty$, where the rate function $I(a)$ is given by the Legendre-Fenchel transform of the scaled cumulant generating function (SCGF) $\lambda(k)$, and that furthermore, $\lambda(k)$ is given by the logarithm of the dominant (largest) eigenvalue of the $M \times M$ matrix $\Pi_{xy}e^{kg\left(x\right)}$.

It is now not difficult to adapt this formalism to sequences $\vect{x}_1, \dots, \vect{x}_N$ generated by deterministic, chaotic maps.
Beginning from the Markov-chain formalism, we first take the continuum limit in order to allow for an infinite, continuous state space, so that the tilted operator becomes
\be
L_{k}\rho\left(\vect{y}\right)=\int\Pi\left(\vect{x},\vect{y}\right)e^{kg\left(\vect{x}\right)}\rho\left(\vect{x}\right)d\vect{x}.
\ee
Now we notice that for a choatic map, the transition probabilities are given by
\be
\Pi\left(\vect{x},\vect{y}\right)=\delta\left(\vect{y}-f\left(\vect{x}\right)\right) \,,
\ee
leading to Eq.~\eqref{Lkdef} of the main text.

\subsection{Reverse process}

To make the paper self contained we provide here more details on the derivation of the reverse process on which our biased MC simulations rely.
The joint PDF of a sequence generated by a map $f(\vect{x})$ is given by
\be
\label{Pforwards}
P_{\text{joint}}\left(\vect{x}_{1},\dots,\vect{x}_{N}\right)=p_{s}\left(\vect{x}_{1}\right)\prod_{i=1}^{N-1}\delta\left(\vect{x}_{i+1}-f\left(\vect{x}_{i}\right)\right) \, .
\ee
This can be rewritten as 
\be
\label{Pbackwards}
P_{\text{joint}}\left(\vect{x}_{1},\dots,\vect{x}_{N}\right)=p_{s}\left(\vect{x}_{N}\right)\prod_{i=1}^{N-1}p\left(\vect{x}_{i}|\vect{x}_{i+1}\right)
\ee
with conditional probabilities
\be
p\left(\vect{x}_{i}|\vect{x}_{i+1}\right)=\sum_{\vect{z}\in f^{-1}\left(\vect{x}_{i+1}\right)}\frac{p_{s}\left(\vect{z}\right)\delta\left(\vect{x}_{i}-\vect{z}\right)}{p_{s}\left(\vect{x}_{i+1}\right) \left|J_{f}\left(\vect{z}\right)\right|} \, .
\ee
The representation \eqref{Pbackwards} is obtained by using
\be
p_{s}\left(\vect{x}_{i}\right)\delta\left(\vect{x}_{i+1}-f\left(\vect{x}_{i}\right)\right)=p_{s}\left(\vect{x}_{i}\right)\sum_{\vect{z}\in f^{-1}\left(\vect{x}_{i+1}\right)}\frac{\delta\left(\vect{x}_{i}-\vect{z}\right)}{ \left|J_{f}\left(\vect{z}\right)\right|} =p_{s}\left(\vect{x}_{i+1}\right)p\left(\vect{x}_{i}|\vect{x}_{i+1}\right)
\ee
in Eq.~\eqref{Pforwards} sequentially, for $i=1,\dots,N-1$.
Note that $p\left(\vect{x}_{i}|\vect{x}_{i+1}\right)$ is normalized, 
\be
\int  p\left(\vect{x}_{i}|\vect{x}_{i+1}\right)d\vect{x}_{i}=  \sum_{\vect{z}\in f^{-1}\left(\vect{x}_{i+1}\right)}\frac{p_{s}\left(\vect{z}\right)}{p_{s}\!\left(\vect{x}_{i+1}\right) \left|J_{f}\left(\vect{z}\right)\right|}=1.
\ee
The last equality follows from the fact that  $p_s(\vect{x})$ satisfies the Frobenius-Perron equation $Lp_s(\vect{x}) = p_s(\vect{x})$.
Eq.~\eqref{Pbackwards} presents an alternative, statistically equivalent 
method for generating random sequence realizations in the form of a Markov chain, as described in the main text.

%

\subsection{Connection between the doubling map and the binomial distribution}

Here we explain the coincidence of the rate function \eqref{IaDoubling} in the main text that we obtained for the doubling map with that of a binomial distribution. 
To that end, we calculate the rate function by using an alternative method, thereby providing a useful check of Eq.~\eqref{IaDoubling} in the main text and, more generally, of the validity of the theoretical framework that we use.

Let us write the binary representation of $x_1$
\be
x_{1}=\sum_{j=1}^{\infty}\eta_{j}2^{-j}
\ee
As is well known, the doubling map shifts the binary representation, so that in fact
\be
x_{i}=\sum_{j=1}^{\infty}\eta_{i+j-1}2^{-j}
\ee
for all $i=1,2,\dots$.
One then finds that 
\bea
\label{DoublingBinary}
A&=&\frac{1}{N}\sum_{i=1}^{N}x_{i}=\frac{1}{N}\sum_{i=1}^{N}\sum_{j=1}^{\infty}\eta_{i+j-1}2^{-j}\underbrace{=}_{m=i+j-1}\frac{1}{N}\sum_{i=1}^{N}\sum_{m=i}^{\infty}\eta_{m}2^{-\left(m-i+1\right)} \nn\\
&=& \frac{1}{N}\sum_{m=1}^{N}\sum_{i=1}^{m}\eta_{m}2^{-\left(m-i+1\right)}+O\left(\frac{1}{N}\right)=\frac{1}{N}\sum_{m=1}^{N}\eta_{m}+O\left(\frac{1}{N}\right) \, .
\eea
On the other hand, since $x_1$ is generated from a uniform distribution on the interval $[0,1]$, it follows that the $\eta_i$'s are independent and identically distributed Bernoulli random variables with $\mathbb{P}\left(\eta_{i}=0\right)=\mathbb{P}\left(\eta_{i}=1\right)=1/2$. In fact, $\eta_i = \xi_i$ where $\xi_i$ was defined in the main text, so  $A = B/N+O\left(1/N\right)$, and since $B/N$ obeys an LDP [descrbied by Eq.~\eqref{DVScaling} of the main text] with the rate function \eqref{IaDoubling} of the main text (see, e.g., \cite{MS2017SM}), so does $A$.

This connection enables us to characterize the set $S_a$ of initial conditions $x_1$ that lead to a given value $A = a$, in the limit $N \to \infty$, as defined in Eq.~\eqref{Sa} of the main text. Using Eq.~\eqref{DoublingBinary}, we find that
\be
\lim_{N\to\infty}\frac{1}{N}\sum_{n=0}^{N-1}g\left(f^{\left(n\right)}\left(x_{1}\right)\right)=\lim_{N\to\infty}\frac{1}{N}\sum_{n=0}^{N-1}\eta_{j}
\ee
(where, as above, $\eta_1,\eta_2,\dots$ is the binary representation of $x_1$.)
Therefore, the set $S_a$ is precisely the set of numbers on the interval $[0,1]$ whose binary representation has a ratio of $1-a:a$ between zeros and ones, as stated in the main text just below Eq.~\eqref{Sa} of the main text.

\subsection{Supports of the rate functions for the tent and logistic maps and rate function of the logistic map at the edges of the support}

By performing a numerical minimization we find that for the tent and logistic maps, $m_{N}=\max_{x_{1}\in\left[0,1\right]}\mathcal{A}\left(x_{1}\right)$ are given, for $N=1,\dots,20$, by
\bea
&& \left\{ 1,0.75,0.75,0.719,0.712,0.703,0.699,0.694,0.691,0.689,0.687,0.685,0.684,0.683,0.681,0.680, 0.680,0.679,0.678,0.677\right\}, \nn\\\\
&& \left\{ 1,0.781,0.793,0.773,0.772,0.767,0.765,0.763,0.761,0.760,0.759, 0.758,0.758,0.757,0.757,0.756,0.756,0.756,0.755,0.755\right\}, \nn\\
\eea
respectively.
On the other hand, $m_N \ge x^*$ for any fixed point of the map, $f(x^*) = x^*$.
These observations strongly suggest that $m_{\infty}=2/3$ and $m_{\infty}=3/4$ for the tent and logistic maps,  respectively, where $m_{\infty}=\lim_{N\to\infty}m_{N}$.
Therefore, the probability that $A > m_\infty$ strictly vanishes in the large-$N$ limit, so,
from the scaling \eqref{DVScaling} in the main text, it then follows that the supports of the two rate functions $I(a)$ are $[0,m_\infty]$.
Indeed, we found this to be the case when calculating the rate functions through numerical diagonalizations of the operators $\tilde{L}_k$, see Fig.~\ref{figIaTentLogistic} of the main text.

We now turn to calculate the rate function of the logistic map at the edges of the support. The calculation is very similar to the tent map case which is given in the main text.
If $x_1\sim4^{-N}$, then $A\sim 1/N$ (since $f'(0) = 4$ for the logistic map). 
Now, since $p_s(x \ll 1) \sim 1/\sqrt{x}$, the probability for this is $\sim2^{-N}$. Then, using the scaling \eqref{DVScaling} in the main text, this leads to the bound $I(0) \le \ln 2$.
Similarly, if $\left|x_{1}-3/4\right|\sim2^{-N}$ then (since $f'(3/4) = -2$ for the logistic map) $\left|A-3/4\right|\sim1 / N$, leading to $I(3/4) \le \ln 2$.
In fact, we find that these inequalities are saturated, $I\left(0\right)=I\left(3/4\right)=\ln2$, as described in the main text.

\subsection{Exact solutions of the eigenvalue problems for the tent and logistic maps} 
 
%
 
Here we exactly solve the eigenvalue problems for the tent and logistic maps, Eqs.~\eqref{psiTentEq} and \eqref{psiLogisticEq} of the main text, respectively. The solution is given in the form of a perturbative expansion in $k$ that can be solved exactly, to arbitrary order.
For both maps, the expansion that we use is (as described in the main text)
\be
\label{perturbationExpansion}
\psi\left(x\right)=p_{s}\left(x\right)\left[1+k\psi_{1}\left(x\right)+k^{2}\psi_{2}\left(x\right)+\dots\right],\qquad\lambda\left(k\right)=k\lambda_{1}+k^{2}\lambda_{2}+\dots.
\ee
and, as we show below, $\psi_n(x)$ is given by an $n$th degree polynomial whose coefficients are found by solving a set of linear equations.
Since $\psi(x)$ is only determined up to a multiplicative constant, we choose its normalization to be such that $\psi(x=0)/p_s(x=0) = 1$ for all $k$, implying that the constant terms of all of the $\psi_i(x)$'s vanish.
Let us demonstrate how this procedure works for the tent map, for which the invariant measure is $p_s(x) = 1$.
The first order equation is Eq.~\eqref{psiTentEq1} of the main text, whose solution,
$\psi_{1}\left(x\right)=x,\;\lambda_{1}=1/2$,
is too given in the main text.
From the $O(k^2)$ terms, we get the equation
\be
\frac{x^{2}}{2}+\left[\psi_{1}\left(\frac{x}{2}\right)+\psi_{1}\left(1-\frac{x}{2}\right)\right]\frac{x}{2}+\frac{1}{2}\psi_{2}\left(\frac{x}{2}\right)+\frac{1}{2}\psi_{2}\left(1-\frac{x}{2}\right)=\psi_{2}\left(x\right)+\lambda_{1}\psi_{1}\left(x\right)+\lambda_{2}+\frac{\lambda_{1}^{2}}{2}
\ee
which, after plugging in the solution to the first order in $k$, simplifies to
\be
\label{psiTentEq2}
\frac{1}{2}\left[1+\psi_{2}\left(\frac{x}{2}\right)+\psi_{2}\left(1-\frac{x}{2}\right)\right]=\psi_{2}\left(x\right)+\lambda_{2}+\frac{1}{8} \, .
\ee
The key step now is to plug in an ansatz of a quadratic polynomial $\psi_{2}\left(x\right)=a_{1}x+a_{2}x^{2}$ (note that $\psi_2(x)$ can be shifted by an additive constant, so without loss of generality, we choose the constant term of the polynomial to vanish). Plugging this ansatz into \eqref{psiTentEq2}, we get
\be
\label{linearEqs2}
\frac{3a_{2}-2}{4}x^{2}+\frac{2a_{1}+a_{2}}{2}x+\lambda_{2}+\frac{1-4a_{1}-4a_{2}}{8}=0\,.
\ee
From the coefficients of $x^2, x^1$ and $x^0$ in Eq.~\eqref{linearEqs2}, we obtain three linear equations in the three unknowns $a_1, a_2$ and $\lambda_2$, whose solution is
\be
a_{1}=-\frac{1}{3},\quad a_{2}=\frac{2}{3},\quad \lambda_{2}=\frac{1}{24} \, .
\ee

Let us now explain why this method works to arbitrary order in $k$.
It is convenient to consider the expansion
\be
\label{LambdaExpansion}
\Lambda\left(k\right)=e^{\lambda\left(k\right)}=\Lambda_{0}+\Lambda_{1}k+\Lambda_{2}k^{2}+\dots,
\ee
where $\Lambda_0 = 1$ is the largest eigenvalue of the original Frobenius-Perron operator with no tilt, $k=0$.
By plugging the expansions \eqref{perturbationExpansion} and \eqref{LambdaExpansion} into Eq.~\eqref{psiTentEq} of the main text, and writing the result as a power series in $k$, one obtains
\be
\label{psiCatExpansionEq}
\frac{1}{2}\sum_{n=0}^{\infty}\sum_{m=0}^{\infty}k^{n+m}\frac{x^{n}}{n!}\left[\psi_{m}\left(\frac{x}{2}\right)+\psi_{m}\left(1-\frac{x}{2}\right)\right]=\sum_{n=0}\sum_{m=0}^{\infty}k^{n+m}\Lambda_{n}\psi_{m}\left(x\right) \, .
\ee
Now it is clear that it is consistent to look for a solution in which $\psi_{m}\left(x\right)$ is a polynomial of degree $m$, as the overall coefficient of $k^\ell$ then becomes a polynomial of degree $\ell$ in $x$, on both sides of the equation. Moreover, once $\psi_0(x), \dots , \psi_{m-1}(x)$ and $\Lambda_0, \dots, \Lambda_{m-1}$ have been found, the $O(k^m)$ terms in \eqref{psiCatExpansionEq} yield $m+1$ linear equations in $m+1$ unknowns: the coefficients of the polynomial $\psi_m(x)$ (except for its constant term, which is chosen to be zero) and $\Lambda_m$, which one solves similar to the case $m=2$ above and $m=1$ in the main text. From $\Lambda_1, \dots, \Lambda_m$ one then extracts $\lambda_1, \dots, \lambda_m$ by expanding the logarithm $\lambda(k) = \ln \Lambda(k)$ as a power siereis in $k$.
A very similar argument works for the logistic map too, the only key property needed being that $\left(\frac{1+\sqrt{1-x}}{2}\right)^{m}+\left(\frac{1-\sqrt{1-x}}{2}\right)^{m}$ is a polynomial in $x$ of degree $\le m$.

For the tent map, we worked out the leading  eight orders in this expansion explicitly, yielding
\bea
&& \psi\left(x\right)=1+kx+k^{2}\left(-\frac{x}{3}+\frac{2x^{2}}{3}\right)+k^{3}\left(\frac{x}{9}-\frac{7x^{2}}{18}+\frac{x^{3}}{3}\right)+k^{4}\left(-\frac{x}{54}+\frac{83x^2}{540}-\frac{23}{90}x^{3}+\frac{2}{15}x^{4}\right) \nn \\
&&+k^{5}\left(-\frac{7x}{1620}-\frac{577x^{2}}{16200}+\frac{317x^{3}}{2700}-\frac{82x^{4}}{675}+\frac{2x^{5}}{45}\right) \!  + k^{6}\left(\frac{91x}{16200}-\frac{1933x^{2}}{1134000}-\frac{901x^{3}}{27000}+\frac{17807x^{4}}{283500}-\frac{433x^{5}}{9450}+\frac{4x^{6}}{315}\right) \nn\\
&&+ k^{7}\left(-\frac{10343x}{3402000}+\frac{1716529x^{2}}{238140000}+\frac{9013x^{3}}{5670000}-\frac{101393x^{4}}{4961250}+\frac{8599x^{5}}{330750}-\frac{2857x^{6}}{198450}+\frac{x^{7}}{315}\right) \nn\\
&&+ k^{8}\left(\frac{775759x}{714420000}-\frac{231438377x^{2}}{50009400000}+\frac{5953831x^{3}}{1190700000}+\frac{9045061x^{4}}{4167450000}-\frac{3891347x^{5}}{416745000}+\frac{366901x^{6}}{41674500}-\frac{257x^{7}}{66150}+\frac{2x^{8}}{2835}\right) + \dots,\nn\\\\
&& \lambda\left(k\right)=\frac{k}{2}+\frac{k^{2}}{24}-\frac{k^{3}}{72}+\frac{41k^{4}}{8640}-\frac{91k^{5}}{64800}+\frac{2347k^{6}}{9072000}+\frac{54457k^{7}}{952560000}-\frac{149967403k^{8}}{1600300800000} + \dots.
\eea
The Legendre transform of $\lambda(k)$ then yields
\bea
I(a) &=& 6\left(a-\frac{1}{2}\right)^{2}+24\left(a-\frac{1}{2}\right)^{3}+\frac{588}{5}\left(a-\frac{1}{2}\right)^{4}+\frac{14496}{25}\left(a-\frac{1}{2}\right)^{5}+\frac{2502624}{875}\left(a-\frac{1}{2}\right)^{6}  \nn\\
&+& \frac{426696576}{30625}\left(a-\frac{1}{2}\right)^{7}+\frac{71732367264}{1071875}\left(a-\frac{1}{2}\right)^{8} + \dots \,,
\eea
the first three terms of which are given in Eq.~\eqref{IaTent} of the main text.
Similarly, the leading six orders for the logistic map give
\bea
\psi\left(x\right)	&=&p_{s}\left(x\right)\left[1+kx+k^{2}\left(-\frac{x}{8}+\frac{x^{2}}{2}\right)+k^{3}\left(\frac{x}{32}-\frac{x^{2}}{8}+\frac{x^{3}}{6}\right)+k^{4}\left(\frac{x}{512}+\frac{13x^{2}}{384}-\frac{x^{3}}{16}+\frac{x^{4}}{24}\right)\right. \nn\\
&+&\left.k^{5}\left(-\frac{25x}{6144}+\frac{x^{2}}{1536}+\frac{7x^{3}}{384}-\frac{x^{4}}{48}+\frac{x^{5}}{120}\right)+k^{6}\left(\frac{377x}{245760}-\frac{241x^{2}}{61440}-\frac{x^{3}}{2880}+\frac{5x^{4}}{768}-\frac{x^{5}}{192}+\frac{x^{6}}{720}\right)+\dots\right], \nn\\\\
\lambda\left(k\right)	&=& \frac{k}{2}+\frac{k^{2}}{16}-\frac{k^{3}}{64}+\frac{3k^{4}}{1024}+\frac{k^{5}}{12288}-\frac{115k^{6}}{294912} + \dots,
\eea
the Legendre transform yielding the rate function 
\be
I\left(a\right)=4\left(a-\frac{1}{2}\right)^{2}+8\left(a-\frac{1}{2}\right)^{3}+24\left(a-\frac{1}{2}\right)^{4}+\frac{208}{3}\left(a-\frac{1}{2}\right)^{5}+\frac{1856}{9}\left(a-\frac{1}{2}\right)^{6}+\dots,
\ee
the first 3 terms of which are given in Eq.~\eqref{IaLogistic} of the main text.


\subsection{Asymptotic behaviors of the rate functions for the tent and logistic maps near the edges of their supports}

We find the asymptotic behaviors of the rate functions for the tent and logistic maps near the edges of their supports by solving the eigenvalue problems, Eqs.~\eqref{psiTentEq} and \eqref{psiLogisticEq} of the main text respectively, in the limits $k \to \pm \infty$.
Let us begin from the tent map, and consider the limit $a \to 0^+$, which corresponds to $k \to -\infty$.
At all $x$ except for a boundary layer at $x\simeq1$, the second term on the left hand side of Eq.~\eqref{psiTentEq} of the main text
is negligible. The equation becomes 
$e^{kx}\psi\left(x/2\right)/2=e^{\lambda\left(k\right)}\psi\left(x\right)$
which yields the leading-order solution $\psi\left(x\right)\simeq e^{2kx}$,  $e^{\lambda\left(k\right)}\simeq1/2$.
This leading-order approximation of $\psi(x)$ turns out to be sufficient for calculating the subleading correction of $\lambda(k)$, as we now show.
Plugging $x=0$ and $ x=1$ into
Eq.~\eqref{psiTentEq} of the main text yields (respectively)
\bea
\label{x0Tent}
\frac{1}{2}\left[\psi\left(0\right)+\psi\left(1\right)\right]	&=&e^{\lambda\left(k\right)}\psi\left(0\right) , \\
\label{x1Tent}
e^{k}\psi\left(\frac{1}{2}\right)	&=& e^{\lambda\left(k\right)}\psi\left(1\right).
\eea
Plugging Eq.~\eqref{x1Tent} into \eqref{x0Tent} and simplifying, we get
\be
\label{psi0andHalfTent}
e^{k}e^{-\lambda\left(k\right)}\psi\left(\frac{1}{2}\right)=\left[2e^{\lambda\left(k\right)}-1\right]\psi\left(0\right) \, .
\ee
Now using $\psi\left(x\right)\simeq e^{2kx}$, we find
$e^{\lambda\left(k\right)}\simeq 1/2+e^{2k}$,
and after applying the Legendre transform to $\lambda(k)$, we obtain
\be
I(a \ll1) \simeq \frac{a}{2}\ln\frac{a}{4}-\frac{a}{2}+\ln2 \, .
\ee
In fact, one can additionally find the leading-order solution for $\psi(x)$ that also correctly describes boundary layer near $x=1$, $\psi\left(x\right)\simeq e^{2kx}\left[1+e^{2k\left(1-x\right)}\right]$.
Now let us turn to the limit $k \to \infty$ for the tent map. This time, for all $x$ except for a boundary layer at $x\simeq1$, it is the first term in
Eq.~\eqref{psiTentEq} of the main text
that is negligible so it becomes $e^{kx}\psi\left(1-x/2\right) / 2\simeq e^{\lambda\left(k\right)}\psi\left(x\right)$, yielding the leading-order solution 
$\psi\left(x\right)\simeq e^{2kx/3}$, $e^{\lambda\left(k\right)}\simeq e^{2k/3} / 2$.
On the other hand, plugging the fixed point $x=2/3$ into
Eq.~\eqref{psiTentEq} of the main text
we get, after simplifying,
\be
\frac{e^{2k/3}}{2}\psi\left(\frac{1}{3}\right)=\left(e^{\lambda\left(k\right)}-\frac{e^{2k/3}}{2}\right)\psi\left(\frac{2}{3}\right) \, .
\ee
Plugging $\psi\left(x\right)\simeq e^{2kx/3}$ into this equation, we get
$e^{\lambda\left(k\right)}\simeq\left(e^{2k/3}+e^{4k/9}\right)/2$, whose Legendre transform yields
\be
I\left(a\right)\simeq\left(3-\frac{9a}{2}\right)\left[\ln\left(3-\frac{9a}{2}\right)-1\right]+\ln2,\quad\frac{2}{3}-a\ll1 \, .
\ee

For the logistic map, the idea behind the calculation is similar as in the tent map.
We begin by plugging
$\psi\left(x\right)=p_{s}\left(x\right)\eta\left(x\right)$ into 
Eq.~\eqref{psiLogisticEq} of the main text
where we recall 
\be
p_{s}\left(x\right)=\frac{1}{\pi\sqrt{x\left(1-x\right)}}
\ee
is the invariant measure. This results in the more convenient equation
\be
\label{etaLogisticEq}
\frac{e^{kx}}{2}\left[\eta\left(\frac{1-\sqrt{1-x}}{2}\right)+\eta\left(\frac{1+\sqrt{1-x}}{2}\right)\right] = e^{\lambda\left(k\right)}\eta\left(x\right)
\ee
for $\eta(x)$.
In the limit $k \to -\infty$, one can neglect the second term on the left hand side of \eqref{etaLogisticEq} (except for a narrow boundary layer at $x \simeq 1$),
leading to the simpler equation
\be
\frac{e^{kx}}{2}\eta\left(\frac{1-\sqrt{1-x}}{2}\right) \simeq e^{\lambda\left(k\right)}\eta\left(x\right)
\ee
which, in turn, yields the leading-order solution
\be
\label{etaSol1}
\eta\left(x\right)\simeq\exp\left[k\sum_{n=0}^{\infty}f_{<}^{-n}\left(x\right)\right]
\ee
and $e^{\lambda\left(k\right)}\simeq1/2$,
where 
\be
f_{<}^{-1}\left(x\right)=\frac{1-\sqrt{1-x}}{2}
\ee
is the smaller of the two preimages of $x$, and $f_{<}^{-n}$ is the $n$th functional power of $f_{<}^{-1}$.
Following the same steps as we did to reach Eq.~\eqref{psi0andHalfTent} for the tent map, one plugs $x=0$ and $x=1$ into Eq.~\eqref{etaLogisticEq} yielding
\be
\label{eta0andHalfLogistic}
e^{-\lambda\left(k\right)}e^{k}\eta\left(\frac{1}{2}\right)=\left[2e^{\lambda\left(k\right)}-1\right]\eta\left(0\right) \, .
\ee
Now, plugging Eq.~\eqref{etaSol1} into \eqref{eta0andHalfLogistic}, we obtain
\be
e^{\lambda\left(k\right)}\simeq\frac{1}{2}+e^{c_{1}k}, \quad c_{1}=1 + \sum_{n=0}^{\infty}f_{<}^{-n}\left(\frac{1}{2}\right) = 1.6973249\dots,
\ee
which, after applying the Legendre transform to $\lambda(k)$, yields the asymptotic behavior
\be
I\left(a\ll1\right) \simeq \ln2+\frac{a}{c_{1}}\left(\ln\frac{a}{2c_{1}}-1\right) \, .
\ee
Finally, for $k \to \infty$, for all $x$ except for a boundary layer at $x\simeq1$, the first term on the left hand side of \eqref{etaLogisticEq} is negligible so the equation becomes
\be
\frac{e^{kx}}{2}\eta\left(\frac{1+\sqrt{1-x}}{2}\right) \simeq e^{\lambda\left(k\right)}\eta\left(x\right)\simeq\frac{e^{3k/4}}{2}\eta\left(x\right),
\ee
which yields the leading-order solution
\be
\label{etaSol2}
\eta\left(x\right) \simeq \exp\left\{ k\sum_{n=0}^{\infty}\left[f_{>}^{-n}\left(x\right)-\frac{3}{4}\right]\right\}
\ee
and $e^{\lambda\left(k\right)}\simeq e^{3k/4}/2$.
Plugging $x=3/4$, the nontrivial fixed point of $f(x)$, into Eq.~\eqref{etaLogisticEq}, we find
\be
\label{etaThreeQuartersLogistic}
e^{3k/4}\eta\left(\frac{1}{4}\right)=\left(2e^{\lambda\left(k\right)}-e^{3k/4}\right)\eta\left(\frac{3}{4}\right) \, .
\ee
Now, plugging Eq.~\eqref{etaSol2} into \eqref{etaThreeQuartersLogistic}, we obtain
\be
e^{\lambda\left(k\right)}\simeq\frac{e^{3k/4}+e^{c_{2}k}}{2}, \quad 
c_{2}=\frac{3}{4}+\sum_{n=0}^{\infty}\left[f_{>}^{-n}\left(\frac{1}{4}\right)-\frac{3}{4}\right]=0.3472134339\dots.
\ee
Applying the Legendre transform to $\lambda(k)$,  we find the asymptotic behavior
\be
I(a) \simeq \ln2+\frac{1}{\tilde{c}_{2}}\left(a-\frac{3}{4}\right)\left\{ \ln\left[\frac{1}{\tilde{c}_{2}}\left(a-\frac{3}{4}\right)\right]-1\right\}, \qquad \frac{3}{4} - a \ll 1,
\ee
where $\tilde{c}_{2} = c_2 - 3/4 = -0.402786566\dots$.

To summarize the results of this section, we have, for the tent and logistic maps,
\bea
I_{\text{tent}}\left(a\right)&\simeq&\begin{cases}
\frac{a}{2}\ln\frac{a}{4}-\frac{a}{2}+\ln2\,, & a\ll1\,,\\[3mm]
\left(3-\frac{9a}{2}\right)\left[\ln\left(3-\frac{9a}{2}\right)-1\right]+\ln2\,, & \frac{2}{3}-a\ll1\,,
\end{cases}\\[3mm]
I_{\text{logistic}}\left(a\right)&\simeq&\begin{cases}
\ln2+\frac{a}{c_{1}}\left(\ln\frac{a}{2c_{1}}-1\right)\,, & a\ll1\,,\\[3mm]
\ln2+\frac{1}{\tilde{c}_{2}}\left(a-\frac{3}{4}\right)\left\{ \ln\left[\frac{1}{\tilde{c}_{2}}\left(a-\frac{3}{4}\right)\right]-1\right\} \,, & \frac{3}{4}-a\ll1,
\end{cases}
\eea
respectively. These asymptotic behaviors are plotted in Fig.~\ref{figIaTentLogistic} of the main text, displaying excellent agreement with MC simulations and with the exact $I(a)$.

\subsection{Monte-Carlo simulations}
Here we give some additional details regarding the Monte-Carlo (MC) simulations that we used in order to produce the data plotted in Figs.~\ref{figIaTentLogistic} and \ref{figIaCat} 
of the main text.
For the tent and logistic maps, we used the biased MC algorithm described in the text by numerically calculating the conditional distributions $P\left(A = a \,|\, B=b\right)$ with ${10}^4$ realizations for each $b\in\left\{ 0,1,\dots,50\right\}$. As described in the main text, the distribution of $A$ is then calculated using Eq.~\eqref{totalProbability} of the main text.

In all of the cases studied in the text, typical fluctuations of $P(A)$ follow a Gaussian distribution in the large-$N$ limit, as described by the quadratic approximation of the rate function around its minimum at $a=1/2$:
\be
P\left(A=a\right)\simeq\frac{1}{\sqrt{2\pi V}}\exp\left[-\frac{1}{2}I''\left(\frac{1}{2}\right)N\left(a-\frac{1}{2}\right)^{2}\right]
\ee 
where $V=1/\left[NI''\left(1/2\right)\right]$ is the variance.
In Figs.~\ref{figIaTentLogistic} (a) and (b) and Fig.~\ref{figIaCat} (a), therefore, the markers are	
\be
I_{N}\left(a\right) \equiv -\frac{1}{N}\ln\left(\sqrt{\frac{2\pi}{NI''\left(1/2\right)}} \, P\left(A=a\right)\right)
\ee
where $P\left(A=a\right)$ is computed from the simulations, and for $I''\left(1/2\right)$, we plugged in its theoretical value from Eqs.~\eqref{IaTent}, \eqref{IaLogistic} and \eqref{IaCat} of the main text, respectively.
Finally, in Fig.~\ref{figIaCat} (b), the markers correspond to the derivative of $dI_{N}/da$ of the numerically computed $I_{N}\left(a\right)$.

\end{widetext}


\begin{thebibliography} {99}

\bibitem{Geisel82} T. Geisel and J. Nierwetberg, \textit{Onset of Diffusion and Universal Scaling in Chaotic Systems}, {\href{https://journals.aps.org/prl/abstract/10.1103/PhysRevLett.48.7}{Phys. Rev. Lett. \textbf{48}, 7 (1982)}}. 
\bibitem{Schell82} M. Schell, S. Fraser, and R. Kapral, \textit{Diffusive dynamics in systems with translational symmetry: A one-dimensional-map model}, {\href{https://journals.aps.org/pra/abstract/10.1103/PhysRevA.26.504}{Phys. Rev. A \textbf{26}, 504 (1982)}}.
\bibitem{Fujisaka82} H. Fujisaka and S. Grossmann, \textit{Chaos-induced diffusion in nonlinear discrete dynamics}, {\href{https://link.springer.com/article/10.1007/BF01420589}{Z. Phys. B \textbf{48}, 261 (1982)}}. 
\bibitem{Sato19} Y. Sato and R. Klages, \textit{Anomalous Diffusion in Random Dynamical Systems},
{\href{https://journals.aps.org/prl/abstract/10.1103/PhysRevLett.122.174101}{Phys. Rev. Lett. \textbf{122}, 174101 (2019)}}. 
\bibitem{AMR22}
T. Albers, D. Müller-Bender, G. Radons, \textit{Anti-persistent random walks in time-delayed systems}, {\href{https://journals.aps.org/pre/abstract/10.1103/PhysRevE.105.064212}{Phys. Rev. E \textbf{105}, 064212 (2022)}}. 




\bibitem{WB16} J. Wouters and F. Bouchet, \textit{Rare event computation in deterministic chaotic systems using genealogical particle analysis}, {\href{https://iopscience.iop.org/article/10.1088/1751-8113/49/37/374002/meta}{J. Phys. A: Math. Theor. \textbf{49}, 374002 (2016)}}.


\bibitem{RWB17} F. Ragone, J. Wouters and F. Bouchet, \textit{Computation of extreme heat waves in climate models using a large deviation algorithm}, {\href{https://www.pnas.org/doi/full/10.1073/pnas.1712645115}{PNAS \textbf{115}, 24 (2018)}}.

\bibitem{RB21} F. Ragone and F. Bouchet, \textit{Rare Event Algorithm Study of Extreme Warm Summers and Heatwaves Over Europe}, {\href{https://agupubs.onlinelibrary.wiley.com/doi/10.1029/2020GL091197}{Geophys. Res. Lett. \textbf{48}, e2020GL091197 (2021)}}.


\bibitem{chaosWeather21} B.-W. Shen, R. A. Pielke Sr., X. Zeng, J.-J. Baik, S. Faghih-Naini, J. Cui, and R. Atlas, \textit{Is Weather Chaotic?: Coexistence of Chaos and Order within a Generalized Lorenz Model}, {\href{https://journals.ametsoc.org/view/journals/bams/102/1/BAMS-D-19-0165.1.xml}{Bull Am Meteorol Soc. \textbf{102}, 148 (2021)}}.

\bibitem{GLRW21} V. M. Gálfi, V. Lucarini, F. Ragone and J. Wouters, \textit{Applications of large deviation theory in geophysical fluid dynamics and climate science}, {\href{https://link.springer.com/article/10.1007/s40766-021-00020-z}{La Rivista del Nuovo Cimento \textbf{44}, 291 (2021)}}. 

\bibitem{GL21} V. M. Galfi and V. Lucarini, \textit{Fingerprinting Heatwaves and Cold Spells and Assessing Their Response to Climate Change Using Large Deviation Theory}, {\href{https://journals.aps.org/prl/abstract/10.1103/PhysRevLett.127.058701}{Phys. Rev. Lett. \textbf{127}, 058701 (2021)}}.


\bibitem{EPYTNLF22} M. d'Errico, F. Pons, P. Yiou, S. Tao, C. Nardini, F. Lunkeit, D. Faranda, \textit{Present and future synoptic circulation patterns associated with cold and snowy spells over Italy}, {\href{https://esd.copernicus.org/articles/13/961/2022/}{Earth Syst. Dynam. \textbf{13}, 961 (2022)}}.




\bibitem{KSB17} I. Klioutchnikova, M. Sigovaa, N. Beizerov, \textit{Chaos Theory in Finance}, {\href{https://www.sciencedirect.com/science/article/pii/S1877050917324067?via\%3Dihub}{Procedia Comput. Sci.
 \textbf{119}, 368 (2017)}}. 

\bibitem{STSAH02} L. A. Safonova, E. Tomer, V. V. Strygin, Y. Ashkenazy, S. Havlin, \textit{Multifractal chaotic attractors in a system of delay-differential equations
modeling road traffic}, {\href{https://aip.scitation.org/doi/10.1063/1.1507903}{Chaos \textbf{12}, 1006 (2002)}}.

\bibitem{FBBC19} F. Roy,  G. Biroli, G. Bunin and C. Cammarota, \textit{Numerical implementation of dynamical mean field theory for disordered systems: application to the Lotka–Volterra model of ecosystems}, {\href{https://iopscience.iop.org/article/10.1088/1751-8121/ab1f32}{J. Phys. A: Math. Theor. \textbf{52}, 484001 (2019)}}.

\bibitem{FBBB20} F. Roy, M. Barbier, G. Biroli, G. Bunin, \textit{Complex interactions can create persistent fluctuations in high-diversity ecosystems}, {\href{https://journals.plos.org/ploscompbiol/article?id=10.1371/journal.pcbi.1007827}{PLoS Comput. Biol. \textbf{16}, e1007827 (2020)}}.

\bibitem{PAF20} M. T. Pearce, A. Agarwala and D. S. Fisher, \textit{Stabilization of extensive fine-scale diversity by ecologically driven spatiotemporal chaos}, {\href{https://www.pnas.org/doi/full/10.1073/pnas.1915313117}{PNAS \textbf{117}, 14572 (2020)}}.

\bibitem{PAT21} O. Postavaru, S. R. Anton and A. Toma, \textit{COVID-19 pandemic and chaos theory}, 
{\href{https://www.sciencedirect.com/science/article/pii/S0378475420303396?via\%3Dihub}{Math. Comput. Simul. \textbf{181}, 138 (2021)}}.

\bibitem{Varadhan} S. S. Varadhan, \textit{Large Deviations and Applications}, CBMS-NSF Regional Conference Series in Applied Mathematics, No. 46 (SIAM, Philadelphia, 1984).
\bibitem{O1989} Y. Oono, Prog.
Theor. Phys. Suppl. \textbf{99}, 165 (1989).
\bibitem{DZ} A. Dembo and O. Zeitouni, \textit{Large Deviations Techniques
and Applications}, 2nd ed. (Springer, New York, 1998).
\bibitem{Hollander} F. den Hollander, \textit{Large Deviations}, Fields Institute Monographs, vol. 14 (AMS, Providence, Rhode Island, 2000).
\bibitem{Bray} S. N. Majumdar and A. J. Bray, \textit{Large-deviation functions for nonlinear functionals of a Gaussian stationary Markov process}, 
{\href {https://journals.aps.org/pre/abstract/10.1103/PhysRevE.65.051112}{Phys. Rev. E \textbf{65}, 051112 (2002).}}

\bibitem{bucklew2004} J. A. Bucklew, \textit{Introduction to Rare Event Simulation}, Springer, New York (2004).

\bibitem{Majumdar2007} S. N. Majumdar, \textit{Brownian Functionals in Physics and Computer Science}, 
{\href{https://arxiv.org/abs/cond-mat/0510064} {Current Science, vol-89, 2076 (2005).}}


\bibitem{Derrida2007} B. Derrida, \textit{Non-equilibrium steady states: fluctuations and large deviations of the density and of the current}, {\href{https://iopscience.iop.org/article/10.1088/1742-5468/2007/07/P07023}{J. Stat. Mech. (2007) P07023.}}

\bibitem{T2009} H. Touchette, \textit{The large deviation approach to statistical mechanics}, {\href {https://www.sciencedirect.com/science/article/pii/S0370157309001410?via\%3Dihub}{Phys. Rep. \textbf{478}, 1 (2009).}}

\bibitem{Waclaw2010} B. Waclaw, R. J. Allen, and M. R. Evans, \textit{Dynamical phase transition in a model for evolution with migration}, {\href{https://journals.aps.org/prl/abstract/10.1103/PhysRevLett.105.268101}{Phys. Rev. Lett. \textbf{105}, 268101 (2010)}}. 

\bibitem{Derrida11} B. Derrida, \textit{Microscopic versus macroscopic approaches to non-equilibrium systems}, {\href{https://iopscience.iop.org/article/10.1088/1742-5468/2011/01/P01030/meta}{J. Stat. Mech. (2011) P01030}}.

\bibitem{touchette2011} H. Touchette, \textit{A basic introduction to large deviations: Theory, applications, simulations}, in: R. Leidl, A.K. Hartmann (Eds.), Modern Computational Science 11: Lecture Notes from the 3rd International Oldenburg Summer School, BIS-Verlag der Carl von Ossietzky Universität Oldenburg, (2011), arXiv:1106.4146.

\bibitem{CohenMukamel2012} O. Cohen and D. Mukamel, \textit{Phase Diagram and Density Large Deviations of a Nonconserving $ABC$ Model}, {\href{https://journals.aps.org/prl/abstract/10.1103/PhysRevLett.108.060602}{Phys. Rev. Lett. \textbf{108}, 060602 (2012)}}.

\bibitem{EvansMajumdar2014} J. Szavits-Nossan, M. R. Evans, and S. N. Majumdar, \textit{Constraint-Driven Condensation in Large Fluctuations of Linear Statistics},
{\href{https://journals.aps.org/prl/abstract/10.1103/PhysRevLett.112.020602}{Phys. Rev. Lett. \textbf{112}, 020602 (2014)}}. 

\bibitem{bertini2015} L. Bertini, A. De Sole, D. Gabrielli, G. Jona-Lasinio, and C. Landim, \textit{Macroscopic fluctuation theory}, {\href{https://journals.aps.org/rmp/abstract/10.1103/RevModPhys.87.593}{Rev. Mod. Phys. \textbf{87}, 593 (2015)}}.

\bibitem{RednerMeerson} B. Meerson and S. Redner, \textit{Mortality, redundancy, and diversity in stochastic search}, {\href{https://journals.aps.org/prl/abstract/10.1103/PhysRevLett.114.198101}{Phys. Rev. Lett. \textbf{114}, 198101 (2015)}}. 

\bibitem{Vivo2015} P. Vivo, \textit{Large deviations of the maximum of independent and identically distributed random variables}, {\href{https://iopscience.iop.org/article/10.1088/0143-0807/36/5/055037}{Eur. J. Phys. \textbf{36}, 055037 (2015)}}. 

\bibitem{Baek15}   Y. Baek and Y. Kafri, \textit{Singularities in large deviation functions
}, {\href{https://iopscience.iop.org/article/10.1088/1742-5468/2015/08/P08026}{J. Stat. Mech. (2015) P08026}}.
\bibitem{Baek17}   Y. Baek, Y. Kafri, and V. Lecomte, \textit{Dynamical Symmetry Breaking and Phase Transitions in Driven Diffusive Systems}, {\href{https://journals.aps.org/prl/abstract/10.1103/PhysRevLett.118.030604}{Phys. Rev. Lett. \textbf{118}, 030604 (2017)}}.
\bibitem{Baek18}   Y. Baek, Y. Kafri, and V. Lecomte, \textit{Dynamical phase transitions in the current distribution of driven diffusive channels}, {\href{https://iopscience.iop.org/article/10.1088/1751-8121/aaa8f9/meta} {J. Phys. A: Math. Theor. \textbf{51}, 10500 (2018)}}.

\bibitem{Shpielberg2016}
O. Shpielberg and E. Akkermans, \textit{Le Chatelier Principle for Out-of-Equilibrium and Boundary-Driven Systems: Application to Dynamical Phase Transitions}, {\href{https://journals.aps.org/prl/abstract/10.1103/PhysRevLett.116.240603}{Phys. Rev. Lett. \textbf{116}, 240603 (2016)}}. 

\bibitem{TouchetteMinimalModel} P. Tsobgni Nyawo, H. Touchette, \textit{A minimal model of dynamical phase transition},  {\href{https://iopscience.iop.org/article/10.1209/0295-5075/116/50009}{Europhys. Lett. \textbf{116}, 50009 (2016).}} 

\bibitem{LeDoussal2017} A. Krajenbrink and P. Le Doussal, \textit{Exact short-time height distribution in the one-dimensional Kardar-Parisi-Zhang equation with Brownian initial condition}, {\href{https://journals.aps.org/pre/abstract/10.1103/PhysRevE.96.020102}{Phys. Rev. E \textbf{96}, 020102(R) (2017)}}.

\bibitem{MeersonAssaf2017} M. Assaf and B. Meerson, \textit{WKB theory of large deviations in stochastic populations}, {\href {https://iopscience.iop.org/article/10.1088/1751-8121/aa669a} {J. Phys. A: Math. Theor. \textbf{50}, 263001 (2017). }}
\bibitem{MS2017} S. N. Majumdar and G. Schehr, \textit{Large deviations}, ICTS Newsletter 2017 (Volume 3, Issue 2); arxiv 1711:07571.
\bibitem{Touchette2018} H. Touchette, \textit{Introduction to dynamical large deviations of Markov processes}, {\href{https://doi.org/10.1016/j.physa.2017.10.046}{Physica A \textbf{504}, 5 (2018)}}.

\bibitem{SKM2018} N. R. Smith, A. Kamenev and B. Meerson, \textit{Landau theory of the short-time dynamical phase transitions of the Kardar-Parisi-Zhang interface}, {\href{https://journals.aps.org/pre/abstract/10.1103/PhysRevE.97.042130}{Phys. Rev. E \textbf{97}, 042130 (2018).}} 


\bibitem{Hartmann2002}
A. K. Hartmann, \textit{Sampling rare events: Statistics of local sequence alignments},
{\href{https://journals.aps.org/pre/abstract/10.1103/PhysRevE.65.056102}{Phys. Rev. E 65, 056102 (2002)}}.





\bibitem{Young1990} L.-S. Young, \textit{Some large deviation results for dynamical systems}, {\href{https://doi.org/10.2307/2001318}{Trans. Amer. Math. Soc. \textbf{318}, 525 (1990)}}.
\bibitem{CLMPV02} F. M. Cucchietti, C. H. Lewenkopf, E. R. Mucciolo, H. M. Pastawski, and R. O. Vallejos,
\textit{Measuring the Lyapunov exponent using quantum mechanics}, {\href{https://journals.aps.org/pre/abstract/10.1103/PhysRevE.65.046209}{Phys. Rev. E \textbf{65}, 046209 (2002)}}.
\bibitem{Haller01}
G. Haller, \textit{Distinguished material surfaces and coherent structures in three-dimensional fluid flows},
{\href{https://doi.org/10.1016/S0167-2789(00)00199-8}{Physica D \textbf{149}, 248 (2001)}}.
\bibitem{Haller05}
G. Haller, \textit{An objective definition of a vortex},
{\href{https://www.cambridge.org/core/journals/journal-of-fluid-mechanics/article/an-objective-definition-of-a-vortex/3CD781A3AEC4BA16571CBC2D9B4E973F}{J. Fluid Mech. \textbf{525}, 1 (2005)}}.
\bibitem{EP86} J.-P. Eckmann and I. Procaccia, \textit{Fluctuations of dynamical scaling indices in nonlinear
systems}, {\href{https://journals.aps.org/pra/abstract/10.1103/PhysRevA.34.659}{Phys. Rev. A \textbf{34}, 659 (1986)}}.
\bibitem{PV87} G. Paladin and A. Vulpiani, \textit{Anomalous scaling laws in multifractal objects}, 
{\href{https://doi.org/10.1016/0370-1573(87)90110-4}{Phys. Rep. \textbf{156}, 147 (1987)}}.
\bibitem{Frisch95} U. Frisch, \textit{Turbulence, The Legacy of A.N. Kolmogorov}, Cambridge University Press (1995).
\bibitem{Bec06} J. Bec, \textit{Lyapunov exponents of heavy particles in turbulence}, {\href{https://aip.scitation.org/doi/10.1063/1.2349587}{Phys. Fluids \textbf{18}, 091702 (2006)}}.


\bibitem{AV15} R. Aimino, S. Vaienti, \textit{A Note on the Large Deviations for Piecewise Expanding Multidimensional Maps}, 	
in Nonlinear Dynamics New Directions. Nonlinear Systems and Complexity, edited by H. González-Aguilar and E. Ugalde, Series on Mathematical Method andModeling Vol. 11 (Springer, Cham, 2015), pp. 1–10, arXiv:1110.5488.





\bibitem{LNKT13}
T. Laffargue, K.-D. Nguyen Thu Lam, J. Kurchan and J. Tailleur1, \textit{Large deviations of Lyapunov exponents}, {\href{https://iopscience.iop.org/article/10.1088/1751-8113/46/25/254002}{J. Phys. A: Math. Theor. \textbf{46}, 254002 (2013)}}.
\bibitem{BMV14} L. Biferale, C. Meneveau and R. Verzicco, 
\textit{Deformation statistics of sub-Kolmogorov-scale ellipsoidal neutrally buoyant drops in isotropic turbulence}, {\href{https://www.doi.org/10.1017/jfm.2014.366}{J. Fluid Mech. \textbf{754}, 184 (2014)}}.
\bibitem{LLA14} J. C. Leitão, J. M. Viana Parente Lopes, and E. G. Altmann, \textit{Efficiency of Monte Carlo sampling in chaotic systems}, {\href{https://journals.aps.org/pre/abstract/10.1103/PhysRevE.90.052916}{Phys. Rev. E \textbf{90}, 052916 (2014)}}.
\bibitem{JM15} P. L. Johnson and C. Meneveau, \textit{Large-deviation joint statistics of the finite-time Lyapunov spectrum in isotropic turbulence}, {\href{https://aip.scitation.org/doi/10.1063/1.4928699}{Phys. Fluids \textbf{27}, 085110 (2015)}}.
\bibitem{PLP16} D. Pazó, J. M. López, and A. Politi, \textit{Diverging Fluctuations of the Lyapunov Exponents}, {\href{https://journals.aps.org/prl/abstract/10.1103/PhysRevLett.117.034101}{Phys. Rev. Lett. \textbf{117}, 034101 (2016)}}. 
\bibitem{LLA17}
J. C. Leitão, J. M. Viana Parente Lopes and E. G. Altmann, \textit{Importance sampling of rare events in chaotic systems}, {\href{link.springer.com/article/10.1140/epjb/e2017-80054-3}{Eur. Phys. J. B \textbf{90}, 181 (2017)}}.
\bibitem{DSDLV18} L. De Cruz, S. Schubert, J. Demaeyer, V. Lucarini, and S. Vannitsem, \textit{Exploring the Lyapunov instability properties of high-dimensional atmospheric and climate models}, {\href{https://npg.copernicus.org/articles/25/387/2018/#Ch1.T1}{Nonlin. Processes Geophys. \textbf{25}, 387 (2018)}}.

\bibitem{YYSL21} K. Yoshida, H. Yoshino, A. Shudo, D. Lippolis, \textit{Eigenfunctions of the Perron-Frobenius operator and the finite-time Lyapunov exponents in uniformly hyperbolic area-preserving maps}, 
{\href{https://iopscience.iop.org/article/10.1088/1751-8121/ac02b7}{J. Phys. A: Math. Theor. \textbf{54}, 285701 (2021)}}.


\bibitem{ACV17}
C. Anteneodo, S. Camargo, and R. O. Vallejos, \textit{Importance sampling with imperfect cloning for the computation of generalized Lyapunov exponents}, {\href{https://journals.aps.org/pre/abstract/10.1103/PhysRevE.96.062209}{Phys. Rev. E \textbf{96}, 062209 (2017)}}. 


\bibitem{Grassberger85} P. Grassberger, \textit{Information flow and maximum entropy measures for 1-D maps}, {\href{https://www.sciencedirect.com/science/article/pii/0167278985900958}{Physica D \textbf{14}, 365 (1985)}}.


\bibitem{GBP88} P. Grassberger, R. Badii and A. Politi, \textit{Scaling laws for invariant measures on hyperbolic and nonhyperbolic atractors}, {\href{https://link.springer.com/article/10.1007/BF01015324}{J. Stat. Phys. \textbf{51}, 135 (1988)}}.

\bibitem{ParisiAppendix84} G. Parisi, Appendix, in U. Frisch, Fully developed turbulence and intermittency, in
\textit{Proceedings of International School on Turbulence and Predictability in Geophysical Fluid
Dynamics and Climate Dynamics}, M. Ghil, ed. (North-Holland, 1984). 

\bibitem{BS93} C. Beck and F. Schl\"ogl, \textit{Thermodynamics of Chaotic Systems},
(Cambridge University Press, Cambridge, 1993).

\bibitem{Anteneodo04} C. Anteneodo, \textit{Statistics of finite-time Lyapunov exponents in the Ulam map}, 
{\href{https://journals.aps.org/pre/abstract/10.1103/PhysRevE.69.016207}{Phys. Rev. E \textbf{69}, 016207 (2004)}}.

\bibitem{footnote:PDF} For brevity, we denote the probability density function (PDF) of $A$ by $P(A=a)$.

\bibitem{footnote:mappings} It is well known that some of the chaotic maps considered in this paper are homeomorphic to each other: a sequence $x_1, x_2, \dots$ generated by a chaotic map $x_{i+1} = f(x_i)$, can be transformed to a sequence $\tilde{x}_i=h(x_i)$ that is generated by some other chaotic map, i.e., $\tilde{x}_{i+1} = \tilde{f}(\tilde{x}_i)$. In particular, this is the case for $f(x) = $ the tent map, and $\tilde{f}(\tilde{x})$ = the logistic map, with $h(x) =\sin^{2}\left(\pi x/2\right)$ \cite{tentMapWiki}.
However, note that in the context of the present paper, applying such a transformation $h(x)$ would affect the observable via $g(x)$.

\bibitem{tentMapWiki} \url{https://en.wikipedia.org/wiki/Tent_map}








\bibitem{Ellis} J. G\"{a}rtner, \textit{On Large Deviations from the Invariant Measure}, {\href{https://epubs.siam.org/doi/10.1137/1122003}{Th. Prob. Appl. \textbf{22}, 24 (1977)}}; 
R. S. Ellis, \textit{Large Deviations for a General Class of Random Vectors}, {\href{https://projecteuclid.org/journals/annals-of-probability/volume-12/issue-1/Large-Deviations-for-a-General-Class-of-Random-Vectors/10.1214/aop/1176993370.full}{Ann. Prob. \textbf{12}, 1 (1984)}}.

\bibitem{SM} See Supplemental Material at ..., which also cites Refs.~\cite{DeBacco2016,TsobgniNyawo2016,Whitelam2018,CMT19,Gutierrez2021,Whitelam2021, CVC22, Monthus22}

\bibitem{footnote:Legendre} Since $\lambda(k)$ is differentiable and strictly convex, the Legendre-Fenchel transform reduces to a Legendre transform \cite{LegendreNutshell}, which, for Eq.~\eqref{lambdaDoubling} yields $I(a)$ via $a=\lambda'\left(k\right)=\frac{e^{k}}{1+e^{k}}$, leading to $k=\ln\frac{a}{1-a}$ which is plugged into $I=ka-\lambda\left(k\right)$ to obtain Eq.~\eqref{IaDoubling}.

\bibitem{LegendreNutshell}
H. Touchette, \textit{Legendre-Fenchel transforms in a nutshell},  \url{https://www.ise.ncsu.edu/fuzzy-neural/wp-content/uploads/sites/9/2019/01/or706-LF-transform-1.pdf} (2005).

\bibitem{Feigenbaum78} M. J. Feigenbaum, \textit{Quantitative universality for a class of nonlinear transformations}, {\href{https://link.springer.com/article/10.1007/BF01020332}{J. Stat. Phys. \textbf{19}, 25 (1978)}}.

\bibitem{Ulam47} S. Ulam and J. von Neumann, \textit{On Combination of Stochastic and Deterministic Processes}, Bull. Am. Math. Soc. \textbf{53}, 1120 (1947).

\bibitem{KKS16} S. Klus, P. Koltai, C. Schütte, \textit{On the numerical approximation of the Perron-Frobenius and Koopman operator}, {\href{http://www.aimsciences.org/journals/displayArticlesnew.jsp?paperID=12983}{J. Comput. Dyn. \textbf{3}, 51 (2016)}}.



\bibitem{footnote:Cat} Equivalently, 
one can define $A$ to be the series $A=\left(\sum_{i=1}^{2N}w_{i}\right)/(2N)$, where $\left(w_{1},w_{2}\right)$ are sampled uniformly from the unit square, and 
$w_{i+1}=\left(w_{i}+w_{i-1}\right)\mod1$.

\bibitem{BSTV97} F. Brini, S. Siboni, G. Turchetti and S. Vaienti, \textit{Decay of correlations for the automorphism of the torus $\mathbb{T}^2$}, {\href{https://iopscience.iop.org/article/10.1088/0951-7715/10/5/012}{Nonlinearity \textbf{10}, 1257 (1997)}}.	



\bibitem{footnote:2ndOrder} However, analogous transitions for stochastic systems have been categorized as first order \cite{TouchetteMinimalModel}.



\bibitem{OWY84} E. Ott, W. D. Withers, and J. A. Yorke, \textit{Is the dimension of chaotic attractors invariant under coordinate changes?}, {\href{https://link.springer.com/article/10.1007/BF01012932}{J. Stat. Phys. \textbf{36}, 687 (1984)}}.

\bibitem{KP87} D. Katzen and I. Procaccia, \textit{Phase transitions in the thermodynamic formalism of multifractals}, {\href{https://journals.aps.org/prl/abstract/10.1103/PhysRevLett.58.1169}{Phys. Rev. Lett. \textbf{58}, 1169 (1987)}}.




\bibitem{Frisch85} U. Frisch, \textit{Ou en est la Turbulence Developpée?} {\href{https://iopscience.iop.org/article/10.1088/0031-8949/1985/T9/023}{Phys. Scripta \textbf{T9}, 137 (1985)}}.


\bibitem{BPPV84} R. Benzi, G. Paladin, G. Parisi, and A. Vulpiani, \textit{On the multifractal nature of fully developed turbulence and chaotic systems}, {\href{https://iopscience.iop.org/article/10.1088/0305-4470/17/18/021/meta}{J. Phys. A \textbf{17}, 3521 (1984)}}.

\bibitem{Jensen85} M. H. Jensen, L. P. Kadanoff, A. Libchaber, I. Procaccia, and J. Stavans, \textit{Global Universality at the Onset of Chaos: Results of a Forced Rayleigh-Bénard Experiment}, {\href{https://journals.aps.org/prl/abstract/10.1103/PhysRevLett.55.2798}{Phys. Rev. Lett. \textbf{55}, 2798 (1985)}};
T. C. Halsey, M. H. Jensen, L. P. Kadanoff, I. Procaccia, and B. I. Shraiman, \textit{Fractal measures and their singularities: The characterization of strange sets}, {\href{https://journals.aps.org/pra/abstract/10.1103/PhysRevA.33.1141}{Phys. Rev. A \textbf{33}, 1141 (1986)}}. 




\bibitem{NT18} D. Nickelsen and H. Touchette, \textit{Anomalous Scaling of Dynamical Large Deviations}, 
{\href{https://journals.aps.org/prl/abstract/10.1103/PhysRevLett.121.090602}{Phys. Rev. Lett. \textbf{121}, 090602 (2018)}}.
\bibitem{GM19} G. Gradenigo, and S. N. Majumdar, {\it A First-Order Dynamical Transition in the displacement distribution of a Driven Run-and-Tumble Particle}, {\href{https://iopscience.iop.org/article/10.1088/1742-5468/ab11be}{J. Stat. Mech. 053206 (2019)}}.
\bibitem{MeersonGaussian19} B. Meerson, \textit{Anomalous scaling of dynamical large deviations of stationary Gaussian processes}, {\href{https://journals.aps.org/pre/abstract/10.1103/PhysRevE.100.042135}{Phys. Rev. E \textbf{100}, 042135 (2019)}}.
\bibitem{Jack20}
R. L. Jack and R. J. Harris, \textit{Giant leaps and long excursions: Fluctuation mechanisms in systems with long-range memory}, {\href{https://journals.aps.org/pre/abstract/10.1103/PhysRevE.102.012154}{Phys. Rev. E \textbf{102}, 012154 (2020)}}. 
\bibitem{BKLP20} F. Brosset, T. Klein, A. Lagnoux, and P. Petit,
{\it Probabilistic proofs of large deviation results for sums of semiexponential random variables and explicit rate function at the transition},
arXiv preprint: arXiv:2007.08164.
\bibitem{GIL21} G. Gradenigo, S. Iubini, R. Livi, and S. N. Majumdar, \textit{Localization transition in the discrete nonlinear Schrödinger equation: ensembles inequivalence and negative temperatures},
{\href{https://iopscience.iop.org/article/10.1088/1742-5468/abda26}{J. Stat. Mech. 023201 (2021)}}.
\bibitem{GIL21b} G. Gradenigo, S. Iubini, R. Livi, and S. N. Majumdar, 
\textit{Condensation transition and ensemble inequivalence in the discrete nonlinear Schrödinger equation},
{\href{https://link.springer.com/article/10.1140\%2Fepje\%2Fs10189-021-00046-5}{Eur. Phys. J. E {\bf 44}, 29 (2021)}}.

\bibitem{MLMS21} F. Mori, P. Le Doussal, S. N. Majumdar, and G. Schehr, \textit{Condensation transition in the late-time position of a run-and-tumble particle}, {\href{https://journals.aps.org/pre/abstract/10.1103/PhysRevE.103.062134}{Phys. Rev. E \textbf{103}, 062134 (2021)}}.
\bibitem{MGM21} F.~Mori, G.~Gradenigo and S.~N.~Majumdar, \textit{First-order condensation transition in the position distribution of a run-and-tumble particle in one dimension},
{\href{https://iopscience.iop.org/article/10.1088/1742-5468/ac2899}{J. Stat. Mech. 103208 (2021)}}.
\bibitem{Smith22OU} N. R. Smith, {\it Anomalous scaling and 
first-order dynamical phase transition in large deviations of the Ornstein-Uhlenbeck process},
{\href{https://journals.aps.org/pre/abstract/10.1103/PhysRevE.105.014120}{Phys. Rev. E \textbf{105}, 014120 (2022)}}. 
\bibitem{SM22} N. R. Smith and S. N. Majumdar, \textit{Condensation transition in large deviations of self-similar Gaussian processes with stochastic resetting}, {\href{https://iopscience.iop.org/article/10.1088/1742-5468/ac6f04}{J. Stat. Mech. (2022) 053212}}. 

\bibitem{Krajnik22a}
Ž. Krajnik, E. Ilievski, T. Prosen, \textit{Absence of Normal Fluctuations in an Integrable Magnet}, {\href{https://journals.aps.org/prl/abstract/10.1103/PhysRevLett.128.090604}{Phys. Rev. Lett. \textbf{128}, 090604 (2022)}}. 
\bibitem{Krajnik22b} Ž. Krajnik, J. Schmidt, V. Pasquier, E. Ilievski, T. Prosen, \textit{Exact anomalous current fluctuations in a deterministic interacting model}, {\href{https://journals.aps.org/prl/abstract/10.1103/PhysRevLett.128.160601}{Phys. Rev. Lett. \textbf{128}, 160601 (2022)}}.



\bibitem{DeBacco2016}
C.~{De Bacco}, A.~Guggiola, R.~K{\"{u}}hn, and P.~Paga, \textit{Rare events statistics of random walks on networks: localisation and other dynamical phase transitions}, {\href{https://iopscience.iop.org/article/10.1088/1751-8113/49/18/184003/meta}{J. Phys. A: Math. Theor., \textbf{49}, 184003 (2016)}}.

\bibitem{TsobgniNyawo2016}
P.~Tsobgni Nyawo and H.~Touchette, \textit{Large deviations of the current for driven periodic diffusions}, {\href{https://journals.aps.org/pre/abstract/10.1103/PhysRevE.94.032101}{Phys. Rev. E, \textbf{94}, 032101 (2016)}}.

\bibitem{Whitelam2018}
S.~Whitelam, \textit{Large deviations in the presence of cooperativity and slow dynamics}, {\href{https://journals.aps.org/pre/abstract/10.1103/PhysRevE.97.062109}{Phys. Rev. E, \textbf{97}, 62109 (2018)}}.

\bibitem{CMT19} F. Coghi, J. Morand, H. Touchette, \textit{Large deviations of random walks on random graphs}, {\href{https://journals.aps.org/pre/abstract/10.1103/PhysRevE.99.022137}{Phys. Rev. E \textbf{99}, 022137 (2019)}}.

\bibitem{Gutierrez2021}
R.~Gutierrez and C.~Perez-Espigares, \textit{Generalized optimal paths and weight distributions revealed through the large deviations of random walks on networks}, {\href{https://journals.aps.org/pre/abstract/10.1103/PhysRevE.103.022319}{Phys. Rev. E, \textbf{103}, 022319 (2021)}}.

\bibitem{Whitelam2021}
S.~Whitelam and D.~Jacobson, \textit{Varied phenomenology of models displaying dynamical large-deviation singularities}, {\href{https://journals.aps.org/pre/abstract/10.1103/PhysRevE.103.032152}{Phys. Rev. E,  \textbf{103}, 032152, (2021)}}.
  				
\bibitem{CVC22} G. Carugno, P. Vivo, F. Coghi, \textit{Graph-combinatorial approach for large deviations of Markov chains}, {\href{https://iopscience.iop.org/article/10.1088/1751-8121/ac79e6}{J. Phys. A: Math. Theor. \textbf{55}, 295001 (2022)}}.

\bibitem{Monthus22} C. Monthus, \textit{Markov trajectories : Microcanonical Ensembles based on empirical observables as compared to Canonical Ensembles based on Markov generators}, {\href{https://link.springer.com/article/10.1140/epjb/s10051-022-00386-x}{Eur. Phys. J. B \textbf{95}, 139 (2022)}}.



\end{thebibliography}

\begin{thebibliography}	{}



\bibitem{GBP88s} P. Grassberger, R. Badii and A. Politi, \textit{Scaling laws for invariant measures on hyperbolic and nonhyperbolic atractors}, {\href{https://link.springer.com/article/10.1007/BF01015324}{J. Stat. Phys. \textbf{51}, 135 (1988)}}.

\bibitem{ParisiAppendix84s} G. Parisi, Appendix, in U. Frisch, Fully developed turbulence and intermittency, in
\textit{Proceedings of International School on Turbulence and Predictability in Geophysical Fluid
Dynamics and Climate Dynamics}, M. Ghil, ed. (North-Holland, 1984). 

\bibitem{AV15s} R. Aimino, S. Vaienti, \textit{A Note on the Large Deviations for Piecewise Expanding Multidimensional Maps}, 	
in Nonlinear Dynamics New Directions. Nonlinear Systems and Complexity, edited by H. González-Aguilar and E. Ugalde, Series on Mathematical Method andModeling Vol. 11 (Springer, Cham, 2015), pp. 1–10, arXiv:1110.5488.
		
			

				
\bibitem{DeBacco2016s}
C.~{De Bacco}, A.~Guggiola, R.~K{\"{u}}hn, and P.~Paga, \textit{Rare events statistics of random walks on networks: localisation and other dynamical phase transitions}, {\href{https://iopscience.iop.org/article/10.1088/1751-8113/49/18/184003/meta}{J. Phys. A: Math. Theor., \textbf{49}, 184003 (2016)}}.

\bibitem{TsobgniNyawo2016s}
P.~Tsobgni Nyawo and H.~Touchette, \textit{Large deviations of the current for driven periodic diffusions}, {\href{https://journals.aps.org/pre/abstract/10.1103/PhysRevE.94.032101}{Phys. Rev. E, \textbf{94}, 032101 (2016)}}.

\bibitem{Whitelam2018s}
S.~Whitelam, \textit{Large deviations in the presence of cooperativity and slow dynamics}, {\href{https://journals.aps.org/pre/abstract/10.1103/PhysRevE.97.062109}{Phys. Rev. E, \textbf{97}, 62109 (2018)}}.

\bibitem{CMT19s} F. Coghi, J. Morand, H. Touchette, \textit{Large deviations of random walks on random graphs}, {\href{https://journals.aps.org/pre/abstract/10.1103/PhysRevE.99.022137}{Phys. Rev. E \textbf{99}, 022137 (2019)}}.

\bibitem{Gutierrez2021s}
R.~Gutierrez and C.~Perez-Espigares, \textit{Generalized optimal paths and weight distributions revealed through the large deviations of random walks on networks}, {\href{https://journals.aps.org/pre/abstract/10.1103/PhysRevE.103.022319}{Phys. Rev. E, \textbf{103}, 022319 (2021)}}.

\bibitem{Whitelam2021s}
S.~Whitelam and D.~Jacobson, \textit{Varied phenomenology of models displaying dynamical large-deviation singularities}, {\href{https://journals.aps.org/pre/abstract/10.1103/PhysRevE.103.032152}{Phys. Rev. E,  \textbf{103}, 032152, (2021)}}.
  				
  				
\bibitem{CVC22s} G. Carugno, P. Vivo, F. Coghi, \textit{Graph-combinatorial approach for large deviations of Markov chains}, {\href{https://iopscience.iop.org/article/10.1088/1751-8121/ac79e6}{J. Phys. A: Math. Theor. \textbf{55}, 295001 (2022)}}.

\bibitem{Monthus22s} C. Monthus, \textit{Markov trajectories : Microcanonical Ensembles based on empirical observables as compared to Canonical Ensembles based on Markov generators}, {\href{https://link.springer.com/article/10.1140/epjb/s10051-022-00386-x}{Eur. Phys. J. B \textbf{95}, 139 (2022)}}.
 
 
 \bibitem{GartnerElliss}
J. G\"{a}rtner, \textit{On Large Deviations from the Invariant Measure}, {\href{https://epubs.siam.org/doi/10.1137/1122003}{Th. Prob. Appl. \textbf{22}, 24 (1977)}}; 
R. S. Ellis, \textit{Large Deviations for a General Class of Random Vectors}, {\href{https://projecteuclid.org/journals/annals-of-probability/volume-12/issue-1/Large-Deviations-for-a-General-Class-of-Random-Vectors/10.1214/aop/1176993370.full}{Ann. Prob. \textbf{12}, 1 (1984)}}.
\bibitem{DZs} A. Dembo and O. Zeitouni, \textit{Large Deviations Techniques
and Applications}, 2nd ed. (Springer, New York, 1998).
\bibitem{HollanderSM} F. den Hollander, \textit{Large Deviations}, Fields Institute Monographs, vol. 14 (AMS, Providence, Rhode Island, 2000).
\bibitem{T2009s} H. Touchette, \textit{The large deviation approach to statistical mechanics}, {\href {https://www.sciencedirect.com/science/article/pii/S0370157309001410?via\%3Dihub}{Phys. Rep. \textbf{478}, 1 (2009).}}


\bibitem{MS2017SM} S. N. Majumdar and G. Schehr, ICTS Newsletter 2017 (Volume 3, Issue 2); arxiv 1711:0757.





\end{thebibliography}
\end{document}